\documentstyle[aaspp4,epsfig]{article}

\newcommand{\au}{\,{\rm AU}}
\newcommand{\ms}{\,{\rm m\,s^{-1}}}
\newcommand{\msyr}{\,{\rm m\,s^{-1}\,yr^{-1}}}
\newcommand{\yr}{\,{\rm yr}}
\newcommand{\zmax}{z_{\rm max}}
\newcommand{\mpsini}{M_p\sin i}
\newcommand{\mj}{\,M_J}
\newcommand{\msun}{\,M_\odot}
\newcommand{\be}{\begin{equation}}
\newcommand{\ee}{\end{equation}}
\newcommand{\prot}{P_{\rm rot}}

\begin{document}

\title{The Lick Planet Search : Detectability and Mass Thresholds}

\author{Andrew Cumming\altaffilmark{1,2}, Geoffrey
W. Marcy\altaffilmark{2,3}, \& R. Paul Butler\altaffilmark{4}}

\altaffiltext{1}{Department of Physics, 366 LeConte Hall, University
of California, Berkeley, CA 94720, email: cumming@fire.berkeley.edu}
\altaffiltext{2}{Department of Astronomy, 601 Campbell Hall,
University of California, Berkeley, CA 94720}
\altaffiltext{3}{Department of Physics and Astronomy, San Francisco
State University, San Francisco, CA 94132, email :
gmarcy@stars.sfsu.edu}
\altaffiltext{4}{Anglo-Australian Observatory, PO Box 296, NSW 1710
Epping, Australia, email : paul@aaoepp.aao.gov.au}

\begin{abstract}
We present an analysis of eleven years of precision radial velocity
measurements of 76 nearby solar type stars from the Lick radial
velocity survey. For each star, we report on variability, periodicity
and long term velocity trends. Our sample of stars contains eight
known companions with mass ($\mpsini$) less than 8 Jupiter masses
($M_J$), six of which were discovered at Lick. For the remaining
stars, we place upper limits on the companion mass as a function of
orbital period. For most stars, we can exclude companions with
velocity amplitude $K\gtrsim 20\ms$ at the 99\% level, or
$\mpsini\gtrsim 0.7\mj\,(a/\au)^{1/2}$ for orbital radii $a\lesssim
5\au$.

We examine the implications of our results for the observed
distribution of mass and orbital radius of companions. We show that
the combination of intrinsic stellar variability and measurement
errors most likely explains why all confirmed companions so far have
$K\gtrsim 40\ms$. The finite duration of the observations limits
detection of Jupiter-mass companions to $a\lesssim 3\au$. Thus it
remains possible that the majority of solar type stars harbor
Jupiter-mass companions much like our own, and if so these companions
should be detectable in a few years. It is striking that more massive
companions with $\mpsini>3\mj$ are rare at orbital radii $4$--$6\au$;
we could have detected such objects in $\sim 90\%$ of stars, yet found
none. The observed companions show a ``piling-up'' towards small
orbital radii, and there is a paucity of confirmed and candidate
companions with orbital radii between $\sim 0.2\au$ and $\sim
1\au$. The small number of confirmed companions means that we are not
able to rule out selection effects as the cause of these features.

We show that the traditional method for detecting periodicities, the
Lomb-Scargle periodogram, fails to account for statistical
fluctuations in the mean of a sampled sinusoid, making it non-robust
when the number of observations is small, the sampling is uneven or
for periods comparable to or greater than the duration of the
observations. We adopt a ``floating-mean'' periodogram, in which the
zero-point of the sinusoid is allowed to vary during the fit. We
discuss in detail the normalization of the periodogram, and the
probability distribution of periodogram powers. We stress that the
three different prescriptions in the literature for normalizing the
periodogram are statistically equivalent, and that it is not possible
to write a simple analytic form for the false alarm probability,
making Monte Carlo methods essential.

\end{abstract}

\begin{center}{\bf To appear in The Astrophysical Journal}
\end{center}

\newpage

\section{Introduction}

In the past few years, high precision radial velocity surveys have had
remarkable success in the discovery of planetary-mass companions
around nearby solar-type stars (for reviews, see Marcy \& Butler 1998
and Marcy, Cochran \& Mayor 1999). Searches for companions (Campbell
et al. 1988; McMillan et al. 1994; Mayor \& Queloz 1995; Walker et
al. 1995; Cochran et al. 1997; Noyes et al. 1997; Marcy \& Butler
1992, 1998) have been carried out with Doppler velocity precision
$\approx 10\ms$, although $3\ms$ has been achieved at Lick observatory
for chromospherically quiet stars (Butler et al. 1996). There are now
17 companions known with masses (the observable is $\mpsini$, where
$i$ is the angle of inclination of the orbit with respect to the line
of sight) below 10 Jupiter masses. In total, several hundred stars
have been monitored for timescales of 3 years to more than 11
years. The detections so far suggest that a few percent of solar type
stars harbor companions of a Jupiter mass or more within a few AU.

These objects have raised many questions regarding the distribution of
the mass and orbital radius of planetary-mass companions, and the
relation of these systems to our own solar system and its giant
planets. For example, a surprise was the discovery of Jupiter-mass
companions in close proximity to their host star. Of the 17 companions
within $2.5\au$, 13 have semi-major axis $a<0.5\au$ and five have
$a<0.1\au$. The archetypal example is the companion orbiting 51 Pegasi
(Mayor \& Queloz 1995) which has a mass ($\mpsini$) of 0.44 Jupiter
masses ($M_J$) and an orbital radius $a=0.05\au$, eight times closer
than Mercury's orbit about the Sun. The orbital parameters of the 17
planetary-mass companions are listed in Table \ref{tab:planets}.

Unfortunately, gleaning the true distribution of companions is
complicated by selection effects which favor the detection of massive,
close companions. It is necessary to establish detection thresholds
for searches for planetary-mass companions before the observations can
be fully interpreted. Walker et al. (1995) monitored 21 bright solar
type stars for 12 years. They carried out a detailed statistical
analysis, and from their upper limits could exclude companions with
$\mpsini\sim 1$--$3\mj$ for periods less than the duration of their
observations ($\approx 12$ years). Nelson \& Angel (1998) used a
simple analytic formalism, together with comparisons with real data,
to investigate the dependence of detection thresholds on the number
and duration of observations and the Doppler errors.

Our aim in this paper is to place the confirmed companions from the
Lick radial velocity survey in context by an analysis of the null
detections. The ongoing Lick survey consists of more than eleven years
of precision Doppler velocity measurements of 107 stars (a few of which
have been added or dropped as the survey progressed), and 200 new
stars have recently been added (Fischer et al. 1999). The original
program has so far identified six planetary-mass companions (70 Vir,
47 UMa, 55$\rho$ Cnc, $\tau$ Boo, $v$ And, GL876, they are marked in
Table \ref{tab:planets}), codiscovered the companion to 16 Cyg B, and
confirmed two discoveries made by other groups (51 Peg and $\rho$ CrB,
see Table \ref{tab:planets} for orbital parameters and references).

We search for periodicities and place upper limits using the 
``floating-mean periodogram'', an extension of the well-known
Lomb-Scargle periodogram (Lomb 1976; Scargle 1982) in which we fit
sinusoids to the data, but allow the zero-point of the sinusoid to
``float'' during the fit. This approach was adopted by Walker et
al. (1995), who were interested in obtaining correct upper limits for
periods greater than the duration of the observations. We show here
that allowing the mean to float is crucial to account for statistical
fluctuations in the mean of a sampled sinusoid. The traditional
Lomb-Scargle periodogram fails in precisely the regime where we demand
it be robust, namely when the number of observations is small, the
sampling is uneven or the period of the sinusoid is comparable to or
greater than the duration of the observations. We carefully consider
the correct normalization of the periodogram, an issue which has been
of some debate in the literature (Horne \& Baliunas 1986; Gilliland \&
Baliunas 1987; Schwarzenberg-Czerny 1996).

The plan of the paper is as follows. We begin in \S 2 by describing
the observations and estimating the velocity variability we expect to
see due to measurement error and intrinsic stellar effects. We discuss
in \S 3 our methods for searching for the signatures of companions in
the radial velocity data. We look for variability in excess of our
prediction, long term trends and periodicities. We list those stars
whose data show interesting variability or periodicities which may
indicate the presence of a yet unconfirmed companion. In \S 4, we
place upper limits on the mass of a possible companion as a function
of orbital radius for each star in the sample.  We continue in \S 5
with a discussion of the implications of our results for the
occurrence rate of planetary-mass companions to solar type stars, and
their distribution in mass and orbital radius. We present our
conclusions in \S 6. In Appendix \ref{app:A}, we give a brief
derivation of the periodogram, and in Appendix \ref{app:B}, we discuss
its normalization.

\section{The Observations}

The Lick radial velocity survey is now more than eleven years old
(Marcy \& Butler 1992, 1998). Precise radial velocity measurements
(current precision $\sim 5\ms$, Butler et al. 1996) are made with the
Lick 3m telescope by using an echelle spectrograph and a comparison
iodine reference spectrum. The exposure time is ten minutes for a star
with $V=5$, allowing several observations per year for each star in
the sample. In this paper, we present an analysis of observations of
76 F, G, and K type stars in the original survey\footnote{There have
been observations of 29 M dwarfs as part of the survey, but we
do not include these in our analysis as they are faint ($V>7$) and
suffer from large measurement uncertainties. These stars are part of a
sample being monitored with the Keck telescope. An analysis of their
radial velocity variability will be presented elsewhere.}. A summary
of the observations is given in Table \ref{tab:stars}. For each star,
we give its HR and HD catalog number, spectral type and rotation
period. We list the number and duration of the observations, the
typical internal velocity error and the rms scatter of the data.
Radial velocities are available upon request from G. Marcy.

\subsection{The Distribution of Errors and Intrinsic Stellar Variability\label{sec:errors}}

Two sources of variability can mask velocity variations due to a
companion: measurement uncertainties and intrinsic stellar
variability. In this section, we attempt to quantify these effects.

The uncertainty in the radial velocity measurement $v$ is estimated
for each observation from the dispersion of the velocities measured by
different spectral segments of the spectrometer. An upgrade to the
spectrograph optics and improvements in modeling in November 1994 led
to an increase in the Doppler precision from $\sigma_D\sim
10$--$15\ms$ to $\sigma_D\sim 5\ms$, with $\sigma_D\sim 3\ms$
achievable in the best cases. This improved Doppler precision is
dominated by photon statistics (for a detailed discussion, see Butler
et al. 1996). In this paper, we shall refer to the pre-November 1994
data as ``pre-fix'' and post-November 1994 data as ``post-fix''. Table
\ref{tab:stars} gives the average internal error $<\sigma_D>$ before
and after November 1994 for each star.

Intrinsic stellar variability arises from magnetic activity or
rotation of features across the stellar surface, such as sunspots or
inhomogeneous convective patterns (Saar \& Donahue 1997). Saar, Butler
\& Marcy (1998, hereafter SBM98) used the Lick radial velocity
variations (post-fix data only) to characterize the relationship
between the rotation period, $P_{\rm rot}$, of a star and its
intrinsic velocity variability, $\sigma_V$. They found that the
variability in excess of the internal errors could be explained by
simple models of sunspot rotation and inhomogeneous convective
flows. We use their results to estimate the typical intrinsic
variability associated with such effects as a function of stellar
rotation period\footnote{ At first sight, it may seem circular to use
the work of SBM98 which was based on the Lick data set (the post-fix
data only). However, our approach is to use their results to
characterize the {\it average} variability typical for a star in the
survey with a particular rotation period.}. After inspecting Figure 2
of SBM98, we find $\sigma_V=10\ms(12\ {\rm days}/P_{\rm
rot})^{1.1}$ for G and K type stars and $\sigma_V=10\ms(10\ {\rm
days}/P_{\rm rot})^{1.3}$ for F stars. The rotation period for each
star (taken from Soderblom 1985, Baliunas, Sokoloff \& Soon 1996, or
Fischer 1999, private communication) is given in Table
\ref{tab:stars}.

\begin{figure}
\epsscale{0.6}
\plotone{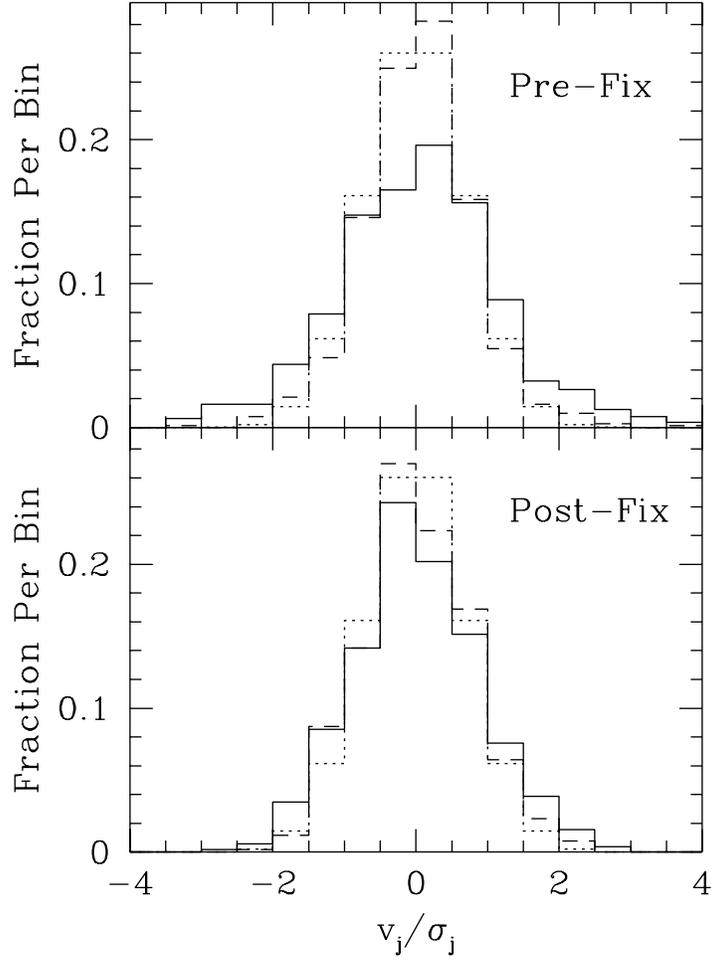}
\caption{ The distribution of Doppler velocities $v_j$ in the
pre-fix and post-fix data. A histogram of the velocities $v_j$ divided
by the estimated error $\sigma_j$ is shown (solid line) for a subset
of 26 stars. These stars, selected from a preliminary analysis of the
{\it post-fix} data, have no significant trend or excess
variability. The number of pre-fix (post-fix) observations is 801
(515). The dotted line in each panel shows a Gaussian distribution
with unit variance. The dashed line in the upper (lower) panel shows
the effect of increasing the pre-fix (post-fix) internal errors by a
constant factor of 1.7 (1.4). These increased internal errors are the
ones we adopt for all stars in our analysis.\label{fig:errdist}}
\end{figure}

We obtain the total estimated variability for each data point by
adding the intrinsic variability to the internal error in quadrature,
$\sigma_j^2=\sigma_V^2+\sigma^2_{D,\,j}$ (here we label each data
point with the index $j$). How well does this estimate reproduce the
scatter in the data?  After a preliminary analysis of the post-fix
data (using the methods of \S\S \ref{sec:trends} and
\ref{sec:variability}), we selected a subset of 26 stars that had no
excess variability or long term trends. In Figure \ref{fig:errdist},
we plot a histogram (solid line) of individual radial velocity
measurements $v_j$ divided by the estimated variability $\sigma_j$ for
this subset of stars. For each star, we have subtracted the weighted
mean of the velocities. The upper panel shows the pre-fix data; the
lower panel shows the post-fix data. The dotted histogram in each case
shows a Gaussian distribution with unit variance. If the scatter in
the velocities were Gaussian with variance $\sigma_j^2$ for each
point, the dotted and solid histograms would match. They do not,
indicating more scatter in the data than we expect. In addition, the
pre-fix and post-fix $v/\sigma$ distributions are different. 

The excess scatter may be due to a combination of underestimated
internal errors and systematic errors in the velocities, particularly
for the early observations. We have chosen to augment the internal
errors by multiplying by a constant factor to force the observed
$v/\sigma$ distribution to match a Gaussian with unit variance. In
this way, {\it we bring the pre-fix and post-fix $v/\sigma$
distributions into agreement}, and we are confident that we have not
underestimated the errors. For all 76 stars in our sample, we multiply
the pre-fix internal errors by 1.7 and the post-fix internal errors by
1.4. The dashed histograms in Figure \ref{fig:errdist} show the
$v/\sigma$ distributions using these rescaled internal errors. The
distribution is unchanged if we remove the 10 chromospherically-active
stars in this subset which have $P_{\rm rot}\leq 12\ {\rm days}$. {\it
Hereafter, we refer to the augmented internal errors as simply
``internal errors''}.

\begin{figure}
\epsscale{0.9}
\plotone{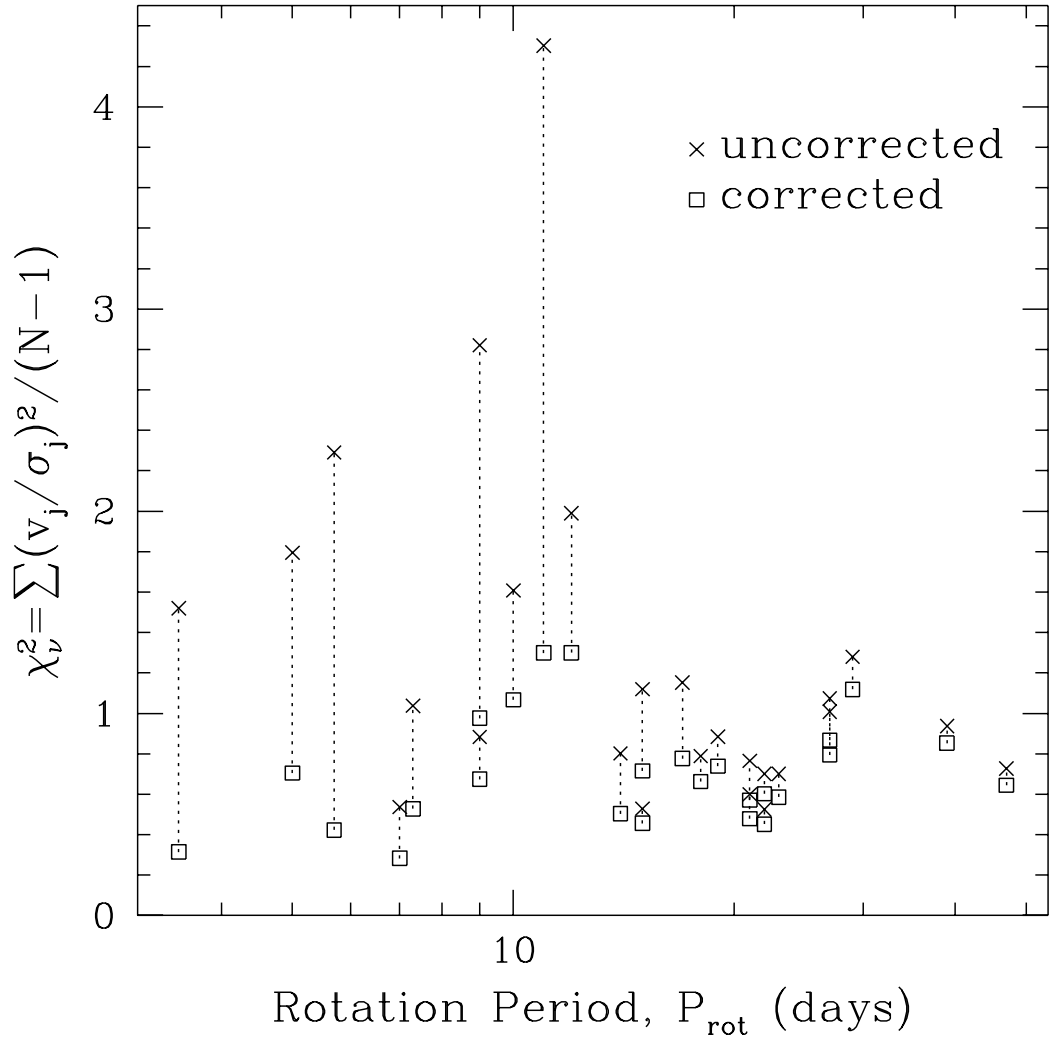}
\caption{ The weighted sum of squares of velocity, $\chi^2=\sum
(v_j/\sigma_j)^2/(N-1)$, as a function of rotation period for the 26
stars of Figure \ref{fig:errdist}. We show the sum evaluated using
internal errors only (crosses) and internal errors plus estimated
intrinsic variability (squares). The intrinsic variability in stars
with $P_{\rm rot}\lesssim 14\ {\rm days}$ is shown by the large
uncorrected $\chi^2$ values for these stars.\label{fig:Prot}}
\end{figure}

In Figure \ref{fig:Prot}, we show the effect of including the
intrinsic variability prediction $\sigma_V$ for stars of different
rotation period.  For each star in the subset of 26 stars of Figure
\ref{fig:errdist}, we plot $\chi^2_\nu=\sum(v_j/\sigma_j)^2/(N-1)$ as
a function of $P_{\rm rot}$. We show $\chi^2_\nu$ evaluated using the
internal errors only (crosses) and including the intrinsic variability
added in quadrature (squares).  The extra variability of stars with
short rotation periods ($P_{\rm rot}\lesssim 12$ days) is shown by the
large uncorrected $\chi^2$ values for these stars. The mean value of
$\chi^2_\nu$ in Figure \ref{fig:Prot} is less than unity, indicating
that our procedure may have overestimated the internal errors
somewhat. However, we prefer to err on the side of {\it
over}estimation.

We use the variability estimate $\sigma_j$ in two ways. The first is
to identify those stars which show much more variability than we might
expect given their rotation periods (\S \ref{sec:variability}). The
second is to weight the data points when we look for periodicities (\S
\ref{sec:periodicity}). The large difference between the pre-fix and
post-fix data makes it important to give less weight to the early data
points. Not only are the pre-fix errors larger, they are less well
characterized than the post-fix errors. One might question the value
of including the low quality early data points at all. However, they
are important because they extend the time baseline, allowing us to
search for longer period companions.

\section{Search for Companions}

The velocity amplitude $K$ of a star of mass $M_{\star}$ due to a
companion with mass $M_p \sin i$ with orbital period $P$ and
eccentricity $e$ is
\be\label{eq:KMp}
K=\left({2\pi G\over P}\right)^{1/3}
{M_p \sin i\over(M_p+M_{\star})^{2/3}}
{1\over \sqrt{1-e^2}}.
\ee
For a circular orbit with $M_p\ll M_\star$, the velocity variations
are sinusoidal with amplitude
\be\label{eq:K}
K=\ 28.4\ {\rm m\,s^{-1}}
\left({1\ {\rm year}\over P}\right)^{1/3}
\left({M_p\sin i\over M_J}\right)
\left({M_\odot\over M_{\star}}\right)^{2/3},
\ee
where $M_J$ is the mass of Jupiter. The orbital period is related to
the orbital radius by Kepler's law,
\be\label{eq:Pa}
P=1\ {\rm year}\ (a/{\rm AU})^{3/2}(M_\odot/M_\star)^{1/2}.
\ee
For example, the companion to 51 Peg ($a=0.05$ AU, $M_p\sin i=0.44$)
induces a velocity amplitude $K=56\ {\rm m\ s^{-1}}$, whereas Jupiter
($a=$5.2 AU, $P=11.9$ years) gives $K=12.5\ {\rm m\,s^{-1}}$ for the
Sun.

In this section, we describe the methods we use to search for such a
signal, and present our results. For each star in Table
\ref{tab:stars}, we first ask if there is a significant long term
trend in the data, and if so we subtract it (\S \ref{sec:trends}). We
then ask if the observed scatter in the data is consistent with the
expected variability (\S \ref{sec:variability}). To search for
periodicities (\S \ref{sec:periodicity}), we fit sinusoids to the
data, employing a generalization of the well-known Lomb-Scargle
periodogram (\S \ref{sec:float}). By using sinusoids as our basic
model for the data, we are strictly assuming circular orbits, although
we find that the periodogram gives a good estimate of the orbital
period even for eccentric orbits. This is because, to lowest order in
the eccentricity, the radial velocity signal from an eccentric orbit
has its main component at the orbital frequency (with smaller components
at multiples of the orbital frequency).  We discuss a possible
extension of the periodogram to Kepler orbits in the conclusions (\S
\ref{sec:concl}). The normalization of the periodogram has been of
some question in the literature, so we discuss our choice of
normalization in Appendix \ref{app:B}. We close this section by
summarizing our results and indicating those stars that show
interesting variability or evidence for companions (\S
\ref{sec:planets}).

\subsection{Long Term Trends}\label{sec:trends}

We first ask if there is a significant {\it long term} trend, on a
timescale much greater than the duration of the observations (i.e.
$\gg$ 10 years). For each star, we fit a straight line $v_j=at_j+b$ to
the measured velocities. When calculating $\chi^2$ for the fit, we
weight each point by the inverse square of the estimated error,
$w_j=1/\sigma_j^2$.  We give the best-fit slope and its uncertainty
(as derived from the least-squares fit) in Table \ref{tab:trends}.

To assess the significance of each slope, we ask if the coefficient of
the linear term is significantly non-zero. We use the F-test to
compare the weighted sum of squares of residuals from the straight
line fit $\chi^2_{N-2}$ (two free parameters) to the weighted sum of
squares about the mean $\chi^2_{N-1}$ (one free parameter). If there is
no long term trend and the residuals are Gaussian distributed, the
statistic
\begin{equation}\label{eq:F}
F=(N-2){\chi^2_{N-1}-\chi^2_{N-2}\over\chi^2_{N-2}}
\end{equation}
follows Fisher's $F$ distribution (Bevington \& Robinson 1992) with
$1$ and $N-2$ degrees of freedom. Given the distribution $F_{1,N-2}$,
we calculate the probability that $F$ would exceed the observed value
$F_{\rm obs}$ purely by chance (the false alarm probability). We give
the F-test false alarm probabilities for each star in Table
\ref{tab:trends}. 

The slopes listed in Table \ref{tab:trends} contain much information
about possible companions at long periods. For the purposes of this
paper, however, we are interested in identifying slopes which would
directly affect our search for companions with $P\lesssim 30$
years. Thus, we mark with $\ast$ in Table \ref{tab:trends} those stars
which have a slope $\geq 5\msyr$ and an F-test false alarm probability
$<10^{-5}$. We adopt a higher threshold ($10^{-5}$) than elsewhere in
this paper because systematic effects in the pre-fix data or
variations due to magnetic activity on long timescales can imitate a
slope. The stars marked with $\ast$ in Table \ref{tab:trends} are HR
219a, HR 753, HR 1325, HR 2047, HR 5544a, HR 5544b, HR 6623, HR 7672
and HR 8086\footnote{HR 8085, the companion to HR 8086, would have
exhibited a slope had we removed the effect of secular acceleration,
see \S \ref{sec:planets}.}.

\subsection{Excess Variability}\label{sec:variability}

We now ask if the scatter in the velocities is consistent with our
predicted variability for each star (\S \ref{sec:errors}). The
$\chi^2$ about the mean, $\chi^2_{N-1}$, or (for those stars marked
$\ast$ in Table \ref{tab:trends}) about the best-fit straight line,
$\chi^2_{N-2}$, gives us a measure of variability. We use the $\chi^2$
distribution with $N-m$ degrees of freedom to test if the velocities
are consistent with being drawn from a Gaussian distribution with
variance $\sigma_j^2$. Here $m$ is the number of parameters in the
model of the data, $m=2$ for a straight line or $m=1$ for a mean. A
small false alarm probability indicates there is more variability in
the data than we expect. The results of this test are shown in Table
\ref{tab:variability}.  We indicate with a $\ast$ those stars which
have false alarm probabilities less than 1\%. We choose a 99\%
threshold so that there will be no more than one false signal in our
sample of 76 stars. The stars which show significant variability are
HR 88, HR 166, HR 2047, HR 4345, HR 5273, HR 5544b, HR 5553, HR 7061,
GL 641, GL 688, and the stars with confirmed planetary-mass companions
(listed in Table \ref{tab:planets}).

\subsection{The Floating-Mean Periodogram}\label{sec:float}

In this section, we test for periodicity in the data using what we
refer to as the ``floating-mean periodogram'', a generalization of the
well-known Lomb-Scargle periodogram (Lomb 1976; Scargle 1982). We
define the floating-mean periodogram as follows. For each trial
frequency $\omega=2\pi/P$, we start with a simple model for the data,
namely a sinusoid plus a constant term,
\begin{equation}\label{eq:model}
f_j=A\cos\omega t_j+B\sin\omega t_j+C,
\end{equation}
where $t_j$ are the observation times. We perform a linear least
squares fit of this model to the data, to determine the constants
$A$,$B$ and $C$. The periodogram is an ``inverted'' plot of the
$\chi^2$ of this fit as a function of frequency. We define the
floating-mean periodogram power $z(\omega)$ as
\begin{equation}\label{eq:float}
z(\omega)={(N-3)\over 2}\ {\chi^2_{N-1}-\chi^2(\omega)\over
\chi^2(\omega_0)},
\end{equation}
where $\chi^2(\omega)=\sum w_j[v_j-f_j(\omega)]^2$ is the $\chi^2$ of
the fit, $\omega_0$ is the best-fit frequency (i.e. the frequency that
gives the maximum periodogram power) and $\chi^2_{N-1}$ is the
weighted sum of squares about the mean. When calculating $\chi^2$, we
weight each data point by $w_j\propto 1/\sigma_j^2$ as in \S
\ref{sec:trends}. We use a linear least squares fitting algorithm from
Press et al. (1992) to fit equation (\ref{eq:model}) to the data and
find $\chi^2(\omega)$. The choice of normalization of the periodogram
has been a subject of some debate in the literature; we discuss this
in detail in Appendix \ref{app:B}. We normalize by the weighted sum of
squares of the residuals to the best fit sinusoid, $\chi^2(\omega_0)$.

The Lomb-Scargle periodogram is obtained by considering a fit of a
sinusoid only, i.e. the case $C\equiv 0$ in equation (\ref{eq:model})
(we sketch the derivation of the Lomb-Scargle formula from the
least-squares approach in Appendix \ref{app:A}). This means that the
zero-point of the sinusoid is assumed to be known already. In
practice, the mean of the data is taken as an estimate of the
zero-point and is subtracted from the data before applying the
Lomb-Scargle formula. In our approach, the zero-point of the sinusoid
is an additional free parameter at each frequency, i.e. the mean of
the data is allowed to ``float'' during the fit.

This approach has been adopted by other authors. Ferraz-Mello (1981)
was the first to do so, defining the ``date-compensated discrete
Fourier transform''. Walker et al. (1995) generalized to the case
where a straight line or quadratic function was subtracted from the
data, defining a ``correlated periodogram''. Most recently, Nelson \&
Angel (1998) included a constant term in their Monte Carlo
experiments. These authors were concerned about the suppression of
periodogram power for periods greater than the duration of the
observations. We show here that allowing the mean to float is
important under much more general circumstances.

\begin{figure}
\epsscale{1.0}
\plotone{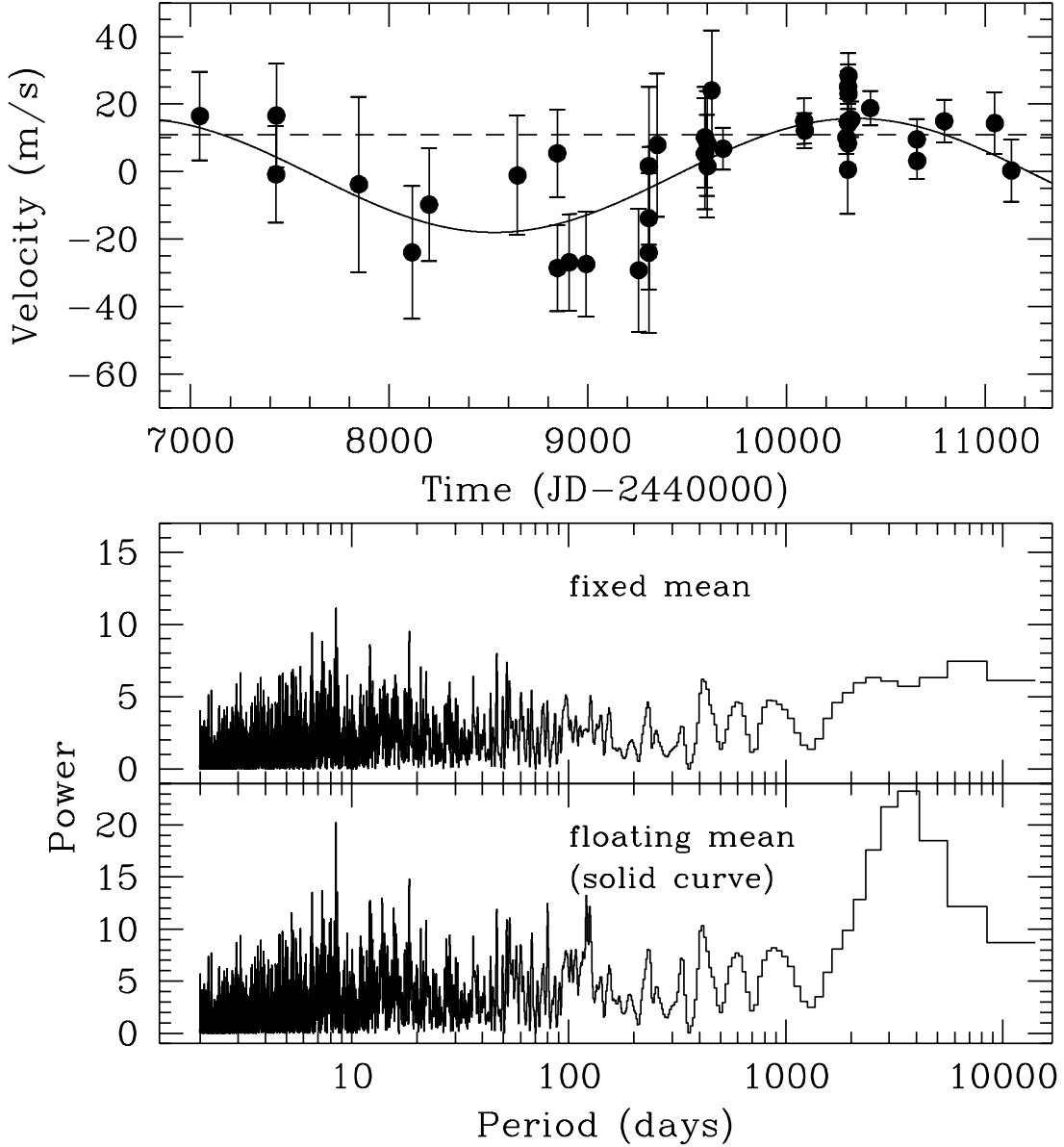}
\caption{An example of the difference between fitting a sinusoid with
the mean of the data subtracted in advance, and fitting a sinusoid
with the mean allowed to ``float'' as an extra free parameter. The
upper panel shows simulated data of a sinusoid with $P=9.6$ years and
amplitude $K=15\ms$. We use the observation times and Doppler errors
for HR 222. The floating-mean periodogram (lower panel) obtains the
correct period ($P=3637$ days), and the rms of the residuals is
$7.7\ms$. The traditional periodogram has a maximum at $P=8.5$
days. The subtraction of the mean results in a suppression of power at
long periods. To allow for a fair comparison, the vertical scales have
a ratio of the square of the rms to account for the different
normalization factors.
\label{fig:222}}
\end{figure}

We now provide some examples which show that allowing the mean to
float is crucial if the number of observations is small, the sampling
is uneven or there is a period comparable to the duration of the
observations or longer. Figure \ref{fig:222} shows simulated data of a
companion in a circular orbit with $P=9.6$ years and $K=15\ms$
($a=4.5\au$, $\mpsini=1.1\mj$). We use the observation times and
errors for one of the stars in our sample, HR 222. The upper panel
shows the velocity measurements as a function of time. The dashed line
shows the mean of the data, which is about $10\ms$ greater than the
correct zero-point of the sine wave. The solid line shows the best-fit
sinusoid when the mean is allowed to float. The traditional and
floating-mean periodograms are shown in the lower panel. The power at
long periods is significantly less in the traditional periodogram than
the floating mean periodogram. Black \& Scargle (1982) first noted
this effect in their analysis of astrometric data, where they showed
that proper motions could significantly reduce the measured amplitude
of long period signals (see Figure 3 of Black \& Scargle 1982).

\begin{figure}
\epsscale{1.0}
\plotone{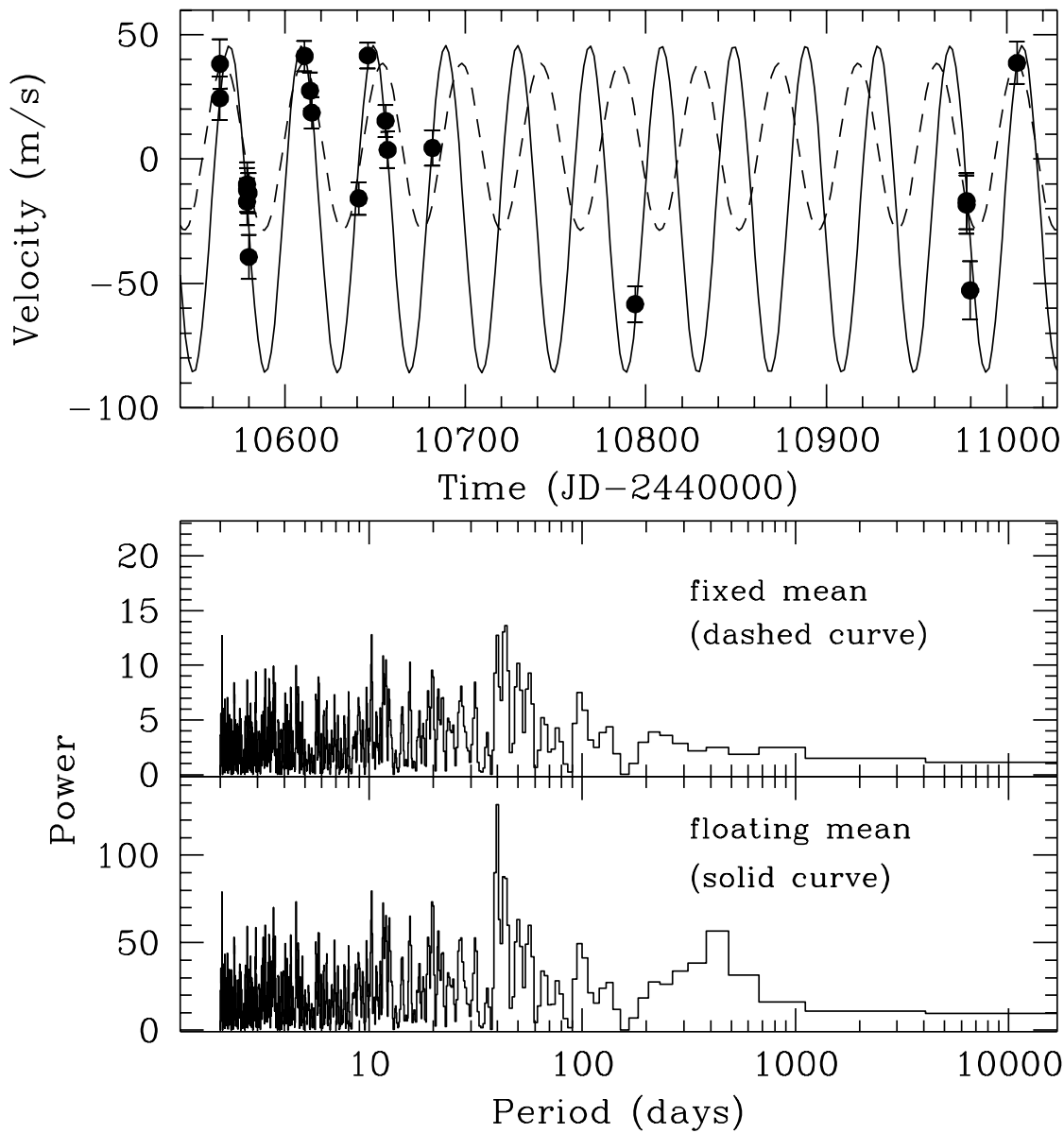}
\caption{A further example of traditional vs. floating-mean
periodogram. The star HR5968 has a companion with $K=67\ms$ and
$P=39.6$ days (Noyes et al. 1997, Table \ref{tab:planets}). The upper
panel shows our 20 radial velocity measurements of HR 5968, spanning
1.2 years. The dashed curve shows the best fit sinusoid ($P=43.9$
days) that we obtain after subtracting the mean of the data. The solid
curve ($P=40.0$ days) is the result of fitting the sinusoid and mean
simultaneously. The rms of the residuals to the dashed curve is
$20\ms$, for the solid curve, it is $8\ms$, exactly what we would
predict from intrinsic variability and measurement errors. The lower
panel shows the periodogram obtained in each case. To allow for a fair
comparison, the vertical scales have a ratio of $(20/8)^2$ to account
for the different normalization factors.
\label{fig:5968}}
\end{figure}

The upper panel in Figure \ref{fig:5968} shows 20 velocities obtained
at Lick for the star HR 5968, which has a known companion with
$K=67\ms$ and $P=39.6$ days (Noyes et al. 1997, Table
\ref{tab:planets}). The duration of these observations is 1.2
years. By chance, most of the measurements lie above the zero-point of
the orbit, giving a $20\ms$ difference between the mean of the data
and the correct zero-point. We plot the best-fit sinusoid with a fixed
mean as a dashed line and with a floating mean as a solid line. The
traditional periodogram does not detect a significant period; it
identifies a period of $43.9$ days, but with false alarm probability
8\%. The floating-mean periodogram gives a very significant detection
at the correct period of $40.0$ days (see Table
\ref{tab:periodogram}). {\it Thus allowing the mean to float is not
only important at long periods, but is crucial to account for
statistical fluctuations when the number of observations is small.} A
similar situation could occur due to uneven sampling. If a sinusoid is
sampled at nearly the same phase each cycle (e.g. for $P\approx 1$
year), the mean of the data could be significantly different from the
zero-point of the sinusoid, and the periodicity thus go
undetected. 

If the number of data points is large and the periodicity is
well-sampled, the mean of the data does give a good estimate of the
correct zero-point. {\it However, the traditional periodogram fails in
precisely the regime where we require it to be robust; namely, when
the number of observations is small, the duration of the observations
is limited, or the sampling uneven.} For this reason, we adopt the
floating-mean periodogram, despite it being less
computationally-efficient than the simple Lomb-Scargle formula.

Following Walker et al. (1995), we make a further generalization in
the case where we subtract a straight line from the data (those stars
marked $\ast$ in Table \ref{tab:trends}). In this case we must fit a
sinusoid plus a straight line to the data at each frequency (i.e. add
a term $\propto t_j$ to eq. [\ref{eq:model}]). For these stars, we use
$\chi^2_{N-2}$ in place of $\chi^2_{N-1}$ in our definition of the
periodogram, and replace $\chi^2(\omega)$ by the $\chi^2$ from the
straight line plus sinusoid fit. This gives a general formula for the
floating-mean periodogram power (as Walker et al. 1995, eq. [A2])
\begin{equation}\label{eq:z}
z(\omega)={(N-m-2)\over 2}\ {\chi^2_{N-m}-\chi^2_{N-m-2}(\omega)\over
\chi^2_{N-m-2}(\omega_0)},
\end{equation}
where $m=1$ for a mean or $m=2$ for a straight-line. Comparing
equation (\ref{eq:z}) with equation (\ref{eq:F}), we see that the
periodogram is similar to the F-statistic, measuring how much the fit
is improved by introducing the two extra sinusoid parameters.

\subsection{Application of the Periodogram}\label{sec:periodicity}

\subsubsection{Search for Periodicities}

For each star, we plot the periodogram and look for the frequency
which gives the maximum power $z_{\rm max}$. The periodogram power $z$
is a continuous function of frequency $f=\omega/2\pi$. However, the
finite duration of the observations $T$ gives each periodogram peak a
finite width $\Delta f\approx 1/T$. Thus in a frequency range $\Delta
f$, there are roughly $T\Delta f$ peaks. To make sure we sample all
the peaks, we evaluate $4T\Delta f$ periods between 2 days and 30
years\footnote{The ``average'' Nyquist period of our observations
($P_{\rm Nyq}\approx 2T/N$) is a few months. However, the uneven
sampling gives information on much shorter periods (perhaps much
shorter than the minimum spacing between observations, see Eyer \&
Bartholdi 1998). The minimum period we investigate is 2 days. The
maximum period of 30 years is a few times greater than the typical
duration of the observations.}. Monte Carlo tests indicate that this
gives adequate sampling of the periodogram. We evaluate $z$ at
evenly-spaced frequencies using a linear least squares fitting
algorithm from Press et al. (1992).

To assess the significance of a possible detection, we test the {\it
null hypothesis} that the data are pure noise. We ask, what is the
false alarm probability associated with $z_{\rm max}$, or how often
would noise fluctuations conspire to give a maximum power larger than
that observed? We use Monte Carlo tests to find the false alarm
probability\footnote{The false alarm probability increases as the
frequency range searched $\Delta f$ increases (Schwarzenberg-Czerny
1996, 1997a, 1998 refers to this as the {\it bandwidth penalty}). As
we discuss in Appendix \ref{app:B}, although the distribution of $z$
at one particular frequency is easy to write down analytically, the
number of independent frequencies in a frequency range $\Delta f$ is
not. For this reason, we adopt a Monte Carlo approach. This also
allows us to check for non-Gaussian effects.}. For each star, we make
fake data sets of either a mean or straight line plus noise. We then
perform the same analysis as for the original data. We find the
maximum periodogram power, and ask, in what fraction of trials does
$z_{\rm max}$ exceeds the observed value? This fraction is the false
alarm probability.

We add noise to the simulated data sets in two ways. One is to add
Gaussian deviates with the same variance as the observed velocities,
keeping the same relative weights and observation times. The second is
to take the observed velocities and randomize them, keeping the sample
times fixed (the so-called bootstrap method, Press et al. 1992). In
this case, we randomize the pre-fix and post-fix velocities
separately, to account for their different $v/\sigma$
distributions. We find that both approaches give false alarm
probabilities which are similar for most stars. We have also applied
the analytic distribution given in Appendix \ref{app:B} to our results
(see Table \ref{tab:app}), and fit for the number of independent
frequencies $M$. The resulting analytic false alarm probabilities
agree well with our Monte Carlo calculations.

The results are given in Table \ref{tab:periodogram}. For each star,
we give the maximum power $z_{\rm max}$, the corresponding best-fit
period and velocity amplitude, and the false alarm probabilities
determined from 400 Monte Carlo trials for each star. We mark with
$\ast$ those stars which show false alarm probabilities less than 1\%.
Apart from those stars with confirmed companions (Table 1), these are
HR 509, HR 937, HR 996, HR 1084, HR 1729, HR 2047, HR 3951, HR 4345,
HR 4496, HR 4540, HR 4983, HR 5019, HR 5273, HR 5544a, HR 5568, HR
6623, HR 7061, HR 7602, HR 7672, and HR 8086. Again, our motivation
for choosing a 99\% detection threshold is that we then expect no more
than one false detection in our sample of 76 stars.

We find that 20 stars have more than one peak in the periodogram with
false alarm probability $<1\%$. The search for multiple companions is
beyond the scope of this paper, however, we carried out a simple test
of whether these secondary peaks were due to aliasing of the primary
period by the finite sampling. We subtracted off the best fit sinusoid
from the data and looked at a periodogram of the residuals to see if
the secondary peaks remained. Only in two cases did they remain, HR
458 which has a second peak at 1210 days, and HR 509 which has a
second peak at 60 days.

\subsubsection{Variability}\label{sec:zav}

The {\it average} power $\bar{z}$ (averaged over frequency) is an
additional indicator of variability. If the data are drawn from a
Gaussian distribution, we expect the mean power to be $\bar{z}\approx
1$ (the mean value of the $F$-distribution). Again using a Monte Carlo
method, we calculate the false alarm probability associated with the
observed $\bar{z}$. Namely, we ask in what fraction of simulated data
sets is the mean power larger than the observed value? A low false
alarm probability in this test indicates either a periodicity may be
present (the uneven sampling results in ``leakage'' which contaminates
the background level), some non-Gaussian behavior or some kind of
``broad-band'' variability (for example as might be expected from
magnetic activity). The results are shown in Table
\ref{tab:zav}. Several stars show significantly high $\bar{z}$ at the
99\% level. Apart from those stars with confirmed companions (Table
\ref{tab:planets}), they are HR 88, HR 166, HR 509, HR 937, HR 996, HR
1084, HR 1729, HR 2047, HR 4345, HR 4496, HR 4540, HR 4983, HR 5019,
HR 5273, HR 5544a, HR 5568, HR 5868, HR 5914, HR 5933, HR 6623, HR
7061, and HR 7602.

\subsection{Results of our Search for Companions}\label{sec:planets}

Our sample of 76 stars contains eight stars with companions of
planetary-mass, HR 3522, HR 4277, HR 458, HR 5072, HR 5185, HR 5968,
HR 7504, and HR 8729. All of these companions give highly significant
periodogram peaks. The orbital period and velocity amplitude obtained
from the periodogram agree well with a Keplerian fit, although not
surprisingly, the velocity amplitude is less than that of the Kepler
fit for eccentric orbits. The orbital parameters for these
planetary-mass companions and references are given in Table
\ref{tab:planets}.

Five stars have companions with $\mpsini>15\mj$ which have been
reported by previous authors. They are HR 2047 ($M\sin i=0.15\msun$,
$P=14.2\yr$; Irwin, Yang, \& Walker 1992), HR 5273 ($M\sin
i=0.4\msun$, $a=5\au$; Kamper 1987), HR 5553 ($K\approx 20\ {\rm km\
s^{-1}}$, $P=125$ d; Beavers \& Salzer 1983), HR 6623 (Cochran \&
Hatzes 1994 report a slope of $-40\pm 5\msyr$) and GL688 ($K=5.7\ {\rm
km\ s^{-1}}$, $P=83.7$ d; Tokovinin 1991). These massive companions
explain the variability seen in these stars, and the trend in those
two, HR 2047 and HR 6623, for which we have not yet seen a complete
orbital period. They provide interesting examples of the limitations
of the periodogram. There are 20 observations of HR 5553. The orbital
period is correctly identified by the periodogram as 125 days, but it
is not deemed a significant detection. GL 688 has only 7
observations, and the periodogram is unable even to identify the
correct period.

Seven stars have significant long-term trends, the slopes of which are
given in Table \ref{tab:summary}. These are HR 219a, HR 753, HR 1325,
HR 5544a, HR 5544b, HR 7672, and HR 8086. Long-term trends such as
these most likely indicate a companion with $a\gtrsim 10\au$ and
$\mpsini>15\mj$. Of course many of the other slopes listed in Table
\ref{tab:trends} may also be due to companions of lower mass (but
still with periods $P\gg 11\yr$). The slopes of HR 5544a and HR 5544b
are of almost the same magnitude but opposite sign, as expected for
orbiting companions. HR 8086 is an interesting case, because much of
its slope is due to secular acceleration, a geometrical effect that
stems from the exchange of proper motion for radial velocity. This
effect is negligible for the other stars in our sample, but we see it
for HR 8086 because it is close and has a large velocity relative to
us. Secular acceleration also explains why we don't see a
corresponding negative slope in HR 8085, the companion to HR 8086.

The 20 remaining stars which show significant variability or
periodicities are listed in Table \ref{tab:summary}. These stars are
candidates for having planetary-mass companions. We include those
stars with a significant periodicity, and those without a significant
periodicity but with variability in excess of our prediction of \S
\ref{sec:errors}. We have divided these stars into two groups,
chromospherically active ($\prot\leq 14$ days) and chromospherically
quiet ($\prot>14$ days). The best fit velocity amplitudes for the
chromospherically quiet stars are of order the predicted scatter in
the velocities due to Doppler errors and intrinsic variability. This
makes it difficult, even for very significant periodogram peaks, to
confidently identify the observed velocity variations with the
Keplerian orbit of a companion. For the chromospherically active stars
($\prot\leq 14$ days), there is an additional complication. Even
though we account for the excess variability expected in
chromospherically active stars, intrinsic velocity variations are
likely to be periodic on many different timescales. This renders
periodicities seen in these stars suspect, as they may be related to
the rotation period, convective motions or magnetic activity. Thus,
despite having extremely low false alarm probabilities in the
periodogram analysis, none of the periodicities listed in Table
\ref{tab:summary} are convincing as companions. We are currently
making more observations of these stars, to attempt to confirm or rule
out these periodicities.

\section{Upper Limits on Companion Mass}\label{sec:limits}

For those stars without a confirmed companion, we would like to know
the upper limit on the signal amplitude $K$. In this section, we use
the periodogram to place upper limits on the velocity amplitude as a
function of period. Strictly, our upper limits are on the amplitude of
{\it circular orbits} because we assume a sinusoidal
periodicity. However, we expect our upper limits would not be
significantly different for eccentric orbits. This is because for a
given period, eccentric orbits have a larger velocity amplitude $K$,
but an extended duration of roughly constant velocity near
apastron. These two effects will cancel each other to some extent for
a companion of given mass and orbital radius. Of course, this is not
true for orbital periods longer than the duration of the observations,
for which our upper limits are strictly for circular orbits. We first
describe our method in detail and then present our upper limits on
companion mass as a function of orbital radius for each star in the
survey.

\subsection{Method}\label{sec:method}

Our method for placing upper limits uses the fact that, because of
measurement errors, different observations of the same signal give
different periodogram powers. Most of the time, a large signal
amplitude will give a large power. Sometimes, because of noise
fluctuations, a large signal will give a small power, but less and
less often as the signal amplitude increases. This means that, given a
particular observation, we can rule out very large signal amplitudes
because they have a very small chance of giving a power as small as
the observed value\footnote{This approach to placing upper limits is a
``frequentist'' one, as discussed by de Jager (1994) (see also Caso et
al. 1998), and applied to the Rayleigh test by Protheroe (1987) and
Brazier (1994). It was used in searches for pulsations in the
quiescent emission from low mass X-ray binaries by Leahy et al. (1983)
and Vaughan et al. (1994). In their work, the noise is dominated by
photon counting (Poisson) noise and the time series is evenly
sampled. This allows the noise level to be ``read off'' the power
spectrum, giving a natural normalization (Leahy et al. 1983). This is
not true in the case of uneven sampling, hence the different
approaches in the literature to normalizing the periodogram (see
Appendix \ref{app:B}).}. We define the 99\% upper limit to be the signal
amplitude such that the periodogram power would be less than or equal
to the observed value only 1\% of the time.

We proceed as follows. For each star, we find the maximum periodogram
power $z_{\rm max}$ from the data (as given in Table
\ref{tab:periodogram}) for periods between 2 days and 30 years. Then
for different trial frequencies, we make simulated data sets of a
sinusoid (with frequency $\omega$, amplitude $K$ and randomly-selected
phase $\phi$) and noise. We find the periodogram power $z(\omega)$ for
each simulated data set and ask how often is $z$ larger than the
observed value $z_{\rm max}$? The 99\% upper limit to the velocity
amplitude at frequency $\omega$ is that $K$ which gives $z>z_{\rm
max}$ in 99\% of trials. In other words, the observed $\zmax$ lies at
the lowest one percentile of the distribution of $z$ that stems from
the upper limit to the velocity.

For each trial, we evaluate the periodogram at only one frequency, the
trial frequency $\omega$. This assumes that the maximum periodogram
power will occur at the trial frequency. We have tested this
assumption by evaluating the upper limits using 100 frequencies
centered on the trial frequency, with the same spacing as used in the
periodogram. The upper limits from both methods agree well, and hence
we adopt the former as it is computationally faster.

To add noise to the simulated data sets, {\it we utilize the residuals
to the sinusoid which best fits the data} (i.e. the residuals after
subtracting from the data the sinusoid corresponding to the maximum
periodogram power $z_{\rm max}$). We motivate this in Appendix
\ref{app:B}, where we show that, if the data consist of a sinusoid
plus noise, the best estimate of the noise variance is the variance of
the residuals to the best fit sinusoid. This allows us to obtain the
noise variance directly from the data, without having to rely on the
estimated errors of \S \ref{sec:errors}. We find that the variance of
the residuals is typically less than the estimated error by a factor
of 1.5, consistent with our finding that we overestimated the internal
errors (see \S \ref{sec:errors}). We are thus confident that the
residuals give a good estimate of the noise variance for each star.

For each observation, we add noise by selecting at random from the
$v/\sigma$ distribution of the residuals, scaling by the expected
variability $\sigma_j$ for each data point. When choosing at random,
we select from the pre-fix and post-fix data separately, as
appropriate. We have also tried adding Gaussian noise (with
appropriate relative weighting between points), scaling by the
variance of the residuals to the best fit sinusoid. The upper limits
from the two methods agree to 10\% or better.

\subsection{Results}\label{sec:upresults}

The 99\% upper limits are plotted in the $\mpsini$--$a$ plane as a
solid line in Figure \ref{fig:allup} for each star in the sample. For
clarity, we state again the meaning of our upper limit. For each star,
we imagine repeating the observations many times, each with identical
sampling, errors and duration as the real observations. The 99\% upper
limit is the mass ($M_p\sin i$) of a companion which, if present at
orbital radius $a$, would give a periodogram power larger than that
observed in the real data ($z_{\rm max}$) in 99\% of our repeated
observations. Thus a companion more massive than the upper limit is
excluded by the data at better than 99\% confidence.

For each star, we calculate the upper limit to the velocity amplitude
at 500 logarithmically-spaced periods between 2 days and 30 years. To
convert velocity and period $(K,P)$ to mass and orbital radius
$(M_p\sin i,a)$, we use equations (\ref{eq:KMp}) and (\ref{eq:Pa}). We
estimate the mass of each star from its spectral type (Table
\ref{tab:stars}) using a simple empirical formula,
\begin{equation}
M/M_{\odot}=1.3-\left({s\over 38}\right)+ {3\over 10}\left(1-{s\over
30}\right)^5,
\end{equation}
where $s$ parameterizes the spectral type, ranging from $s=0$ for F0 to
$s=30$ for M0 (for example, $s=22$ refers to spectral type K2). This
formula reproduces Table 9.6 of Lang (1991) to within
10\%. Metallicity renders our masses uncertain by a further 10\% (see
Carney et al. 1994). This is adequate for our purposes, however.

Inspection of the upper limits shows that, for many periods, the upper
limit to the velocity amplitude $K$ is roughly independent of
period. For each star, we list in Table \ref{tab:K} the average
velocity upper limit $\bar{K}$ calculated for periods less than half
the duration of the observations ($P<T/2$). We show the corresponding
line of constant velocity in Figure \ref{fig:allup} as a dotted
line. The average velocity upper limit is a good estimate of the upper
limit for most periods less than the duration of the observations ($\sim$ 11
years for most stars, see Table \ref{tab:stars}).

The upper limit deviates from this ``constant velocity'' behavior for
two reasons. First, the upper limit is larger at periods where the
sampling of the observations gives poor phase coverage. For example,
many stars show an increase in the upper limit at $a\approx 1\ {\rm
AU}$ because of the tendency for observations to take place at the
same time each year. As expected, these aliasing effects are more
important in stars with fewer observations. Second, the upper limit to
the velocity amplitude increases for periods greater than the duration
of the observations. In Figure \ref{fig:allup}, we show the orbital
radius for which $P=T$ for each star by a vertical dashed line. The
upper limit at long periods ($P\gtrsim T$) is sensitive to how we
choose the phase in the simulated data sets. The data contain some
information about the best fit phase at each period, but we
(conservatively) ignore that and choose the phase at random for each
trial.

In Figure \ref{fig:allup} we do not show results for the 8 stars with
confirmed planetary-mass companions (Table \ref{tab:planets}), or for
the 5 stars with companions with $\mpsini>15\mj$ and $P\gtrsim 20\yr$
(HR2047, HR5273, HR5553, HR6623 and GL 688). An interesting question
is whether these stars have second companions. We are currently
generalizing our approach to the case of two companions. For now, we
note that the residuals to Keplerian fits to the velocities for the
five stars with $\mpsini>15$ are less than $30\ms$ in each
case. Searches for second companions to the stars with planetary-mass
have been made (see Table \ref{tab:planets} for references), without
success except HR 3522 ($\rho^1$ 55 Cnc) and HR 458 (Ups And) show
evidence for long period second companions ($P\gtrsim 3$ years).

Walker et al. (1995) (hereafter W95) also used a floating-mean
periodogram to place upper limits on companion mass, but took a
different approach based on subtracting a sinusoid directly from the
data\footnote{This kind of approach to placing confidence limits is
discussed by Lampton, Margon \& Bowyer (1976), including the cases
where the noise level must be estimated from the data and some of the
parameters are ``uninteresting'' (see also Cline \& Lesser 1970, Avni
1976, Cash 1976 and the discussion in Press et al. 1992).}. We have
applied our method to the data of W95 and find good agreement with
their published upper limits. Comparison with their Figure 5 shows
that our upper limits on companion mass are about 10--20\% lower for
$P<T$ and 20--30\% higher for $P>T$. In addition, our upper limits,
which show less variability from one period to the next, seem less
affected by the sampling of the data. The reason for these differences
is not clear. Our comparison shows that there is a ``theoretical
uncertainty'' in the upper limits of about 20\%. The good agreement of
the two techniques, which are quite different, is encouraging.


\section{Discussion}\label{sec:discuss}

So far, we have searched for companions and determined upper limits on
the mass of companions for each individual star. We now investigate
the implications of our results for the population of planetary-mass
companions as a whole.

The observed distribution of the mass and orbital radius of all known
planetary mass companions is shown in Figure \ref{fig:planets}, in
which we plot the 17 confirmed companions listed in Table
\ref{tab:planets} in the $\mpsini$--$a$ plane. The dashed lines show
constant velocity contours of $10$ and $40\ms$ for a 1 $M_\odot$
star. Four features of the observed distribution are
interesting. First, all confirmed companions have $K\gtrsim
40\ms$. Second, no companions have been detected with orbital radius
$a\gtrsim 2.5\au$. Third, there is a ``piling-up'' of companions at
small orbital radii; for example, of the 17 companions within
$2.5\au$, 13 have semi-major axis $a<0.5\au$ and 5 have
$a<0.1\au$. Fourth, there is a paucity of companions with orbital
radii between $\sim 0.3$ and $\sim 1\au$. In this section, we ask
whether our results help to explain these features.

\subsection{Confirmed and Candidate Companions}

First, we summarize the results of our search for companions. We began
with a sample of 76 stars from the original Lick Survey. Two stars, HR
5968 ($\rho$ CrB) and HR 8729 (51 Peg) were included in the sample
because of the discovery of the companions to these stars by other
groups (Table \ref{tab:planets}), and for this reason cannot be
considered part of a statistically unbiased sample. Of the remaining
74 stars, six have confirmed planetary-mass companions, or about 8\%
of our sample. Our periodogram analysis of \S \ref{sec:periodicity}
reveals several candidate periodicities (marked ``$\ast$'' in Table
\ref{tab:periodogram} and listed in Table \ref{tab:summary}), which
are yet to be confirmed or ruled out as being due to companions.

We plot the confirmed companions (circles) and candidate periodicities
(squares) in the $\mpsini$--$a$ plane in Figure \ref{fig:candid}. We
do this for purely illustrative purposes: {\it we stress that it may
be that none of the candidate periodicities are actually due to
companions.}  Open and filled squares indicate candidate periodicities
for chromospherically active and quiet stars, respectively.  The error
bars show the 99\% upper limit on $\mpsini$ obtained as described in
\S \ref{sec:method}. Broken circles show those confirmed companions
not included in our sample of 74 stars. The dashed lines show constant
velocity contours of 5, 10, 20 and 40$\ms$ for a 1$\ M_\odot$ star.

Inspection of Table \ref{tab:summary} shows that there are several
candidates for companions with velocity amplitude between the Doppler
errors, $K\sim 5$--$10\ms$, and the lowest velocity amplitude
detected, $K\sim 40\ms$. Thus, one interpretation of our results is
that there is an effective detection threshold of $K\approx
40\ms$. This could be caused by the fact that confirmation of these
low amplitude orbits is difficult, as we discussed in \S
\ref{sec:planets}. However, there is an interesting difference in the
velocity amplitudes of the candidate periodicities from
chromospherically quiet versus chromospherically active stars.  Figure
\ref{fig:candid} shows that there are no candidate periodicities with
$K\gtrsim 15\ms$ in the quiet stars, whereas all of the candidate
periodicities in the active stars have $K\gtrsim 15\ms$. Thus another
interpretation is that there is a paucity of companions with velocity
amplitudes $K\sim 15$--$40\ms$, as none are seen in the subset of
quiet stars for which these velocity amplitudes could have been
detected. However, we cannot draw conclusions from the present data,
as the small number of detections is subject to $\sqrt N$
fluctuations.

Figure \ref{fig:candid} also shows that there are fewer candidates
with orbital radii between $a\sim 0.2\au$ and $a\sim 1\au$ than at
other orbital radii. If these periodicities are due to intrinsic
stellar variations, a possible explanation is that these variations
naturally occur on two timescales, the stellar rotation period
($\lesssim$ 1 month) and the timescale of the magnetic cycle
($\gtrsim$ years). However, the apparent paucity of confirmed
companions and candidate periodicities between $a\sim 0.2\au$ and
$a\sim 1\au$ is interesting, and may indicate a real paucity of
companions at these orbital radii.

\subsection{Upper Limits and Detectability}

We now turn to the upper limits which we calculated in \S
\ref{sec:limits}. In Figure \ref{fig:kdist}, we show a histogram (left
panel) and cumulative histogram (right panel) of the mean velocity
upper limits listed in Table \ref{tab:K}. For most stars, we can
exclude companions with $K\gtrsim 10$--$20\ms$ at the 99\% level. Nine
stars (15\% of the sample) have a mean upper limit $\bar{K}<10\ms$ and
38 stars (60\% of the sample) have $\bar{K}<20\ms$. About 10 stars
have $\bar{K}\gtrsim 40\ms$: inspection of Tables \ref{tab:stars} and
\ref{tab:K} shows that these stars have either a small number of
observations, poor internal errors (for example, because they are
faint), short rotation period or a large rms $>40\ms$ (for example,
because of magnetic activity).

In Figure \ref{fig:ncont}, we plot the cumulative histogram in the
$\mpsini$--$a$ plane. We take into account the different stellar
masses and the effects of sampling (i.e. we use the upper limits as a
function of period, Figure \ref{fig:allup}, rather than just
$\bar{K}$, Table \ref{tab:K}). The solid lines show contours of the
number of stars from which a companion of given mass and orbital
radius can be excluded at the 99\% level (plotted on a 40 x 40
grid). Each contour is labeled by the number of stars from which we
can exclude companions above and to the left of the solid line. The
dashed lines show constant velocity contours of 5, 10, 20 and 40$\ms$
for a 1$\ M_\odot$ star. The filled squares show the six confirmed
planetary-mass companions in our sample (Table \ref{tab:planets}). In
Figure \ref{fig:nxb}, we show sections of this contour map. We plot
the number of stars from which a companion can be excluded (at the
99\% level) as a function of orbital radius for different masses, and
as a function of mass for different orbital radii.

Figure \ref{fig:nxb} demonstrates the effect of the finite duration of
the observations. Even for massive companions (e.g. $\mpsini>5\mj$),
the number of stars from which a companion can be excluded decreases
rapidly for $a\gtrsim 6\au$. In Figure \ref{fig:ncont}, we plot with
an open circle the point $\mpsini=1\mj$, $a=5.2\au$, or where our
solar system would lie in this diagram if $\sin i=1$. This shows that
analogs of our solar system are excluded from only a handful of stars
once the distribution of $\sin i$ is taken into account. This is not
true for more massive companions, with $\mpsini>3$--$4\mj$. Our
results show that these companions are not common at orbital radii
$a\gtrsim 1\au$, as found by Walker et al. (1995) in their survey of
21 stars. For example, companions with $\mpsini>3\mj$ ($>5\mj$) can be
excluded from 80\% (95\%) of our sample of 63 stars.

Our results give some sense of how the detectability of a companion of
a particular mass falls off with radius. For example, for
$\mpsini=1\mj$ ($0.5\mj$), the number of stars excluded falls off for
$a\gtrsim 0.3$--$0.5\au$ ($0.1\au$). Thus we have detected all
companions in the sample with $a\lesssim 0.3\au$ and $\mpsini\gtrsim
1\mj$.  For the 40 stars with $\bar{K}\lesssim 20\ms$, we can exclude
$1\mj$ companions out to $1\au$, and $2\mj$ companions out to
$4\au$. Unfortunately, we cannot assume that the detectability of
companions is the same in other surveys. Thus we must restrict our
attention to the six companions in our sample.

We have performed a simple test of whether the six companions in our
sample could have been drawn from a parent population which is
uniformly distributed in orbital radius. First, we assume that the
mass distribution of companions is uniform in $\log\mpsini$ between
0.1 and $4\mj$. This seems consistent with the observed distribution
of $\mpsini$ (at least for small orbital radii where detectability is
good). We select companions at random from this mass distribution and
a uniform distribution in orbital radius from 0 to $2.5\au$. We assign
the companion to a star in the sample at random, simulate observations
of the star using the observation times and errors from the real data,
and ask whether the periodogram power is larger than that
observed. The ``corrected'' orbital radius distribution is then the
distribution of orbital radii of those companions which give
periodogram powers larger than the observed values. We then use the
Kolmogorov-Smirnov (KS) test (Press et al. 1992) to find the
probability that the six companions are drawn from this ``corrected''
distribution.

We find that the probability that the six companions are drawn from a
parent population with a uniform distribution in orbital radius is
$\approx 10\%$. If we do not include 70 Vir, which may have a low
$\sin i$ and thus a mass $>15\mj$, the probability is 5\%. {\it Thus
we cannot rule out that the distribution of companions is uniform in
radius}. Also, we emphasize again that we are dealing with a small
number of detections which are subject to $\sqrt N$
fluctuations. Future detections will enable us to use the KS test in
this way to rule out distributions in mass and orbital radius. For
now, without knowledge of the detection thresholds of other surveys,
we cannot rule out the possibility that the ``pile-up'' of companions
at low orbital radius or the lack of companions between $\sim 0.2\au$
and $\sim 1\au$ are due to selection effects or small number
statistics.

\section{Summary and Conclusions}\label{sec:concl}

We have presented an analysis of eleven years of precision Doppler
velocity measurements of 76 G, K and F type stars from the Lick radial
velocity survey. We have performed tests for variability, long term
trends, and periodicities. Our sample contains eight confirmed
planetary-mass companions, six of which were discovered at Lick, and
five companions of stellar or sub-stellar mass (\S
\ref{sec:planets}). Seven stars have significant long term trends,
likely indicating a companion with $a\gtrsim 10\au$ and
$\mpsini>15\mj$, and 20 stars show variability or periodicities which
may indicate a planetary-mass companion (Table \ref{tab:summary}). We
are currently making more observations of these stars, to attempt to
confirm or rule out these periodicities. For those stars without a
confirmed companion, we have calculated upper limits to the mass of a
companion as a function of orbital radius (Figure
\ref{fig:allup}). For most stars, the mean limit on the velocity
amplitude is between 10 and $20\ms$ or $0.35$ and
$0.7\mj\,(a/\au)^{1/2}$ (for stellar mass $M\approx M_\odot$).

We have searched for periodicities and placed upper limits using a
``floating-mean'' periodogram, in which we fit sinusoids to the data,
allowing the zero-point of the sinusoid to ``float'' during the fit.
Allowing the mean to float is crucial to account for statistical
fluctuations in the mean of a sampled sinusoid. The traditional
Lomb-Scargle periodogram fails when the number of observations is
small, the sampling is uneven or the period of the sinusoid is
comparable to or greater than the duration of the observations. This
may lead to missed detections, or inaccurate upper limits. We have
also expanded on the recent discussion by Schwarzenberg-Czerny (1997a,
1997b, 1998) of the correct way to normalize the periodogram and the
resulting distribution of periodogram powers. The three different
prescriptions in the literature for normalizing the periodogram are
statistically equivalent, and all three give a distribution of
periodogram powers for Gaussian noise which is significantly different
from the usually assumed exponential distribution (see Figure
\ref{fig:dist}). Unfortunately, it is not possible in general to write
a simple analytic formula for the false alarm probability, making
Monte Carlo methods essential.

Our results help to explain the observed distribution of mass and
orbital radius of companions (see Figure \ref{fig:planets}). The
confirmed companions so far have velocity amplitudes $K>40\ms$,
whereas the Doppler errors lie between $5$--$15\ms$. This is most
likely because there is an effective detection threshold which comes
about because of the ambiguity introduced by intrinsic velocity
flutter (which may be periodic). Confirmation of an orbit is then
difficult when the velocity amplitude is similar to the scatter
predicted because of intrinsic variability and Doppler errors. We
note, however, that in the chromospherically quiet stars ($P_{\rm
rot}\gtrsim 14$ days), there are no candidate periodicities in the
range $15$--$40\ms$. This may reflect a real paucity of companions in
this range.

The finite duration of the observations makes it difficult to detect
Jupiter-mass companions with orbital radius $a\gtrsim 3\au$. Thus the
four known companions with $a>1\au$ may be only the first of a
population of Jupiter-mass companions at large orbital radii. This is
not true for more massive companions, however. It is striking that
companions with $\mpsini>3\mj$ are rare at orbital radii $4$--$6\au$;
we could have detected such objects in $\sim 90\%$ of stars, yet found
none.

A few more years of observations will allow detection of Jupiter-mass
companions at $a\sim 5\au$, particularly as the poorer quality pre-fix
observations become less important. Already, we are able to exclude
velocity amplitudes of $10\ms$ from 15\% of the stars in the sample
(for $a\lesssim 3\au$). The velocity amplitudes which can be detected
are ultimately limited by intrinsic stellar variability. Even for
chromospherically inactive stars ($P_{\rm rot}\gtrsim 14$ days), there
is intrinsic flutter of a few m/s. Detectability of short period
companions should improve in the future as there is more feedback
between the observed variability and future observations (ie. stars
which show variability are observed more often). Care must be taken to
include this effect in assessment of detectabilities.

Our analysis has assumed that the orbits of companions are
circular. Yet eccentric orbits appear to be the norm for many of the
planetary-mass companions (see Table \ref{tab:planets} and Marcy et
al. 1999). We find, however, that the periodogram gives a good
estimate of the period for eccentric orbits. In addition, we do not
expect our upper limits to change substantially for eccentric orbits,
as we argued in \S \ref{sec:limits}, except for long period orbits for
which eccentricities are important. The possibility of a non-zero
eccentricity makes identification of an orbital period {\it
impossible} for periods more than two or three times the duration of
the observations. A possible extension of the periodogram to
non-circular orbits would be to fit a Kepler orbit at each frequency,
and define the periodogram power in terms of the $\chi^2$ of the
Kepler fit (see eq. [\ref{eq:z}]). It is not clear if the gain in
detectability would outweigh the computing power needed for this
non-linear least squares fit. A better approach may be to fit higher
harmonics as suggested by Schwarzenberg-Czerny (1996).

The observed distribution of companions in mass and orbital radius
shows a ``piling-up'' towards small orbital radii and a paucity of
companions between orbital radii $a\sim 0.2\au$ and $a\sim
1\au$. Because of the small number of companions in our sample, it is
not possible for us to say whether these features are due to selection
effects. Unfortunately, without knowledge of the detection thresholds
of the other radial velocity surveys, we cannot include the other
confirmed companions in our analysis. The candidate periodicities we
find also show fewer candidates between $a\sim 0.2\au$ and $a\sim
1\au$, which is intriguing. Future detections will show whether these
features reflect the parent population of planetary-mass
companions. It may be that companions are to be found at all orbital
radii, or it may be that there are two populations of Jupiter-mass
companions, one at orbital radii $a\lesssim 0.2\au$ and one at orbital
radii $a\gtrsim 1\au$.

Either scenario presents an interesting challenge to
theorists. Orbital migration models have been proposed to explain the
presence of giant planets at small orbital radii (Lin et al. 1999;
Murray et al. 1998; Trilling et al. 1998; Ward 1997). These models
naturally predict a ``piling-up'' at small orbital radii (Trilling et
al. 1998) because the orbital migration timescale grows progressively
shorter as the planet spirals inwards. However, it is not clear
exactly how such migration might be halted, in particular at orbital
radii as large as $0.2\au$. Indeed, the inevitability of migration may
be responsible for the low percentage of solar type stars which have
close planetary-mass companions (Ward 1997). If migration depends on
gap formation, one would expect migration only to occur for companions
above a certain mass. As yet, there is no observed dependence of the
mass distribution on orbital radius, except for the lack of companions
with $\mpsini\gtrsim 3$--$4\mj$ at large orbital radius, $a\sim
3$--$5\au$. 

There have also been suggestions that gravitational scattering of
planets by other planets, companion stars or neighbouring stars in a
young star cluster may play a role in determining the final
distribution of orbital radii (Rasio \& Ford 1996; Weidenschilling \&
Marzari 1996; Lin \& Ida 1997; Laughlin \& Adams 1998; Levison,
Lissauer \& Duncan 1998). The large range of orbital eccentricities of
the observed companions may be evidence for this type of scenario
(Marcy et al. 1999). One way to lose enough energy to allow a planet
to move from $\sim 5\au$ to $<1\au$ may be interaction with the
protoplanetary disk during the last stages of dissipation, as
suggested by Marcy et al. (1999). It is not known to what extent these
different physical mechanisms play a role in determining the
distribution of planet masses and orbital radii. Clear theoretical
predictions are needed if the discovery of more planetary mass
companions is to allow us to distinguish between these different
pictures.

\acknowledgments We would like to thank Lars Bildsten, Debra Fischer,
Evan Scannapieco, Jonathan Tan and Andrew Youdin for useful
discussions as this work progressed. Debra Fischer provided rotation
periods for several stars, for which a program developed by Phil
Shirts was used to measure the Ca IR triplet emission. Jeff Scargle
and the referee, Andy Nelson, provided thoughtful
comments on the manuscript. Alex Schwarzenberg-Czerny kindly provided
a copy of his paper (Schwarzenberg-Czerny 1998). AC was partly
supported by the Victor F. Lenzen Memorial Scholarship Fund at UC
Berkeley. GWM was partly supported by NSF grant AST 95-20443 and NASA
grant NAGW-3182.

\appendix

\section{Derivation of the Lomb-Scargle Periodogram}\label{app:A}

In this section, we sketch the derivation of the traditional
Lomb-Scargle formula from a least squares fit of a sinusoid to the
data, and show how it relates to the floating mean periodogram used in
this paper. Following Lomb (1976), for each trial frequency
$\omega=2\pi/P$, our model for the velocities is
\begin{equation}
f_j(t_j)=A\cos\omega (t_j-\tau)+B\sin\omega (t_j-\tau).
\end{equation}
The constant $\tau$ is introduced to simplify the calculation of
$\chi^2$, as the cross terms cancel if we choose $\tau$ such that
\begin{equation}\label{eq:tau}
\sum w_j\sin\omega(t_j-\tau)\cos\omega(t_j-\tau)=0,
\end{equation}
or equivalently
\begin{equation}
\tan (2\omega\tau)={\sum w_j\sin 2\omega t_j\over 
\sum w_j\cos 2\omega t_j}.
\end{equation}
The best fit values of the parameters $A$ and $B$ are those which
minimize $\chi^2=\sum w_j (v_j-f_j)^2$, where $w_j\propto
1/\sigma_j^2$ is the weight for data point $j$. Setting
$\partial\chi^2/\partial A=0$ and $\partial\chi^2/\partial B=0$ and
using equation (\ref{eq:tau}) gives
\begin{eqnarray}\label{eq:AB}
A={\sum{w_jv_j}\cos\omega(t_j-\tau)\over\sum w_j\cos^2\omega(t_j-\tau)},
\nonumber\\
B={\sum{w_jv_j}\sin\omega(t_j-\tau)\over\sum w_j\sin^2\omega(t_j-\tau)}.
\end{eqnarray}
The amplitude $K$ and phase $\phi$ of the sinusoid are obtained from
$K^2=A^2+B^2$ and $\tan\phi=A/B$.

Using equation (\ref{eq:AB}) to substitute the best fit $A$ and $B$
into $\chi^2=\sum w_j (v_j-f_j)^2$, we find the minimum value of
$\chi^2$ at each frequency is
\begin{equation}
\chi^2_{\rm min}(\omega)
=\sum w_jv_j^2-\sum w_jf_j^2,
\end{equation}
where $f_j$ is now the best fit sinusoid. As in \S \ref{sec:float}, we
define the unnormalized periodogram power $\hat{z}$ as the reduction
in the sum of squares (Lomb 1976),
\begin{equation}\label{eq:zhat}
\hat{z}(\omega)\equiv\sum w_j f_j^2=\sum w_j v_j^2-\chi^2_{\rm min}(\omega),
\end{equation}
or
\begin{eqnarray}\label{eq:LSapp}
\hat{z}(\omega) = {\left[\sum
w_jv_j\cos\omega(t_j-\tau)\right]^2\over
\sum w_j\cos^2\omega (t_j-\tau)} + \nonumber\\
{\left[\sum w_jv_j\sin\omega(t_j-\tau)\right]^2\over
\sum w_j\sin^2\omega (t_j-\tau)}.
\end{eqnarray}
Equation (\ref{eq:LSapp}) is the (unnormalized) Lomb-Scargle
periodogram (Lomb 1976; Scargle 1982; Horne \& Baliunas 1986) modified
for unequally-weighted data (Gilliland \& Baliunas 1987; Irwin et
al. 1989; Scargle 1989). Equation (\ref{eq:zhat}) shows that the
floating-mean periodogram we adopt in this paper (\S \ref{sec:float})
is a straightforward generalization of the traditional periodogram
(see also Walker et al. 1995, eq. [A2]).

\section{Normalization of the Periodogram and The Distribution of Noise Powers}
\label{app:B}

There are three different prescriptions in the literature for
normalizing the periodogram, namely dividing by (i) the sample
variance (Horne \& Baliunas 1986; Irwin et al. 1989; Walker et
al. 1995), (ii) the variance of the residuals to the best fit sinusoid
(Gilliland \& Baliunas 1987, this paper), or (iii) the variance of the
residuals to the best fit sinusoid {\it at each frequency}
(Schwarzenberg-Czerny 1996). The motivation for these normalizations
is to give the periodogram power $z$ a simple statistical distribution
when the data are pure noise. The goal is to assess the false alarm
probability associated with a given periodogram power. Recent work by
Schwarzenberg-Czerny (1997a, 1997b, 1998) has shown that all three
normalizations are statistically equivalent, and, at a given
frequency, each leads to a simple analytical distribution for Gaussian
noise. We now expand on this discussion and describe our choice of
normalization.

We point out that because of non-orthogonality between different
frequencies, it is difficult to estimate the number of independent
frequencies for a given data set. The commonly used empirical formula
of Horne \& Baliunas (1986) for the number of independent frequencies
is based on an inaccurate probability distribution, and is valid only
for a particular frequency range. We stress that, unfortunately, it is
not possible to write a simple analytic form for the false alarm
probability, making Monte Carlo methods essential.

Why normalize the periodogram at all?  Scargle (1982) and Horne \&
Baliunas (1986) showed that if the data points $X_j$ are independent
Gaussian deviates with variance $\sigma_0^2$, the distribution of
unnormalized periodogram powers is
\begin{equation}
f(\hat{z})\ d\hat{z}=
{1\over\sigma_0^2}\exp(-{\hat{z}\over\sigma_0^2})\ d\hat{z}
\end{equation}
(or simply $\chi^2$ with $2$ degrees of freedom). This analysis
extends to the weighted periodogram, where $\sigma^2_0$ is a measure
of the overall normalization of the weights. We would like to stress
that if the noise variance were somehow known in advance\footnote{This
is often the situation in X-ray astronomy, for example, where the
noise is dominated by Poisson photon statistics, giving a well-defined
background power level (Leahy et al. 1983). See also the discussion in
Lampton, Margon \& Bowyer (1976).}, one could simply normalize by the
known variance $\sigma_0^2$, thus obtaining
\begin{equation}
f(z)\ dz=\exp(-z)\ dz,
\end{equation}
an exponential distribution of periodogram powers. However, in many
cases $\sigma_0^2$ is not known accurately in advance and must be
estimated from the data. The idea is to normalize the noise powers to
a known level\footnote{As we noted in \S \ref{sec:method}, in
principle the background noise level in the power spectrum gives a
direct measure of $\sigma_0^2$. However, in {\it practice} the uneven
sampling results in contamination of the noise powers by the signal
because of spectral leakage and aliasing effects.}, aiding
identification of localized features in the power spectrum
(e.g. periodic signals).

To estimate the noise level from the data set, we use an ``analysis
of variance'' approach (Schwarzenberg-Czerny 1989; Davies 1990). In \S
\ref{app:A}, we showed that
\begin{equation}\label{eq:app1}
\sum w_jv_j^2=\hat{z}(\omega)+\chi^2_{\rm min}(\omega).
\end{equation}
We now define
\begin{equation}
s^2\equiv{1\over N-m}\sum w_jv_j^2,\hspace{1cm}
s_f^2\equiv {\hat{z}\over 2},\hspace{1cm}
s_n^2\equiv {1\over N-m-2}\sum w_j(v_j-f_j)^2,
\end{equation}
and rewrite equation (\ref{eq:app1}) in terms of variances, giving
\begin{equation}\label{eq:app2}
(N-m)s^2=2s_f^2+(N-m-2)s^2_n.
\end{equation}
We have partitioned the variance into two pieces, together with their
respective degrees of freedom; one piece from the signal $s_f^2$ and
one piece from the noise $s_n^2$. Schwarzenberg-Czerny (1989, 1997a,
1998) showed that under the null hypothesis, $s_n^2$ and $s_f^2$ are
statistically independent by Fisher's Lemma. This is also true if
$v_j$ consists of a sinusoidal signal plus Gaussian noise. In this
case, $s_n^2$ is an unbiased estimate of the noise variance. The
different definitions of the normalized periodogram are simply
different ratios of the variances in equation (\ref{eq:app2}). The
``traditional'' normalization by the sample variance is $z\equiv
s_f^2/s^2$ whereas normalizing by the residuals to the noise gives
$z\equiv s_f^2/s_n^2$.

The distribution of powers when the $X_j$ are independent Gaussian
deviates can be written down analytically (Schwarzenberg-Czerny 1997a,
1997b). The partitioning of the degrees of freedom means that $s^2$,
$s_f^2$ and $s_n^2$ are $\chi^2$ distributed with $N-m$, $2$ and
$N-m-2$ degrees of freedom respectively. Thus $z=s_f^2/s_n^2$ follows
an $F$ distribution with $2$ and $N-m-2$ degrees of freedom (for
example, see Abromowitz \& Stegun 1971, \S 26.6). The distribution of
$z=s_f^2/s^2$ is more complex as $s^2$ is correlated\footnote{This
correlation was neglected by Koen (1990), who incorrectly presumed an
$F$ distribution in this case.} with $s_f^2$. The distribution of $z$
in this case is an incomplete beta function (see Abromowitz \& Stegun
1971, \S 26.5). The probability that the periodogram power $z$ is
larger than a given value $z_0$ is given in Table \ref{tab:app} for
the different normalizations.

Given a probability distribution for the periodogram power, we can
write down an expression for the false alarm probability. If the
probability that a periodogram power $z$ is above some value $z_0$ is
${\rm Prob}(z>z_0)$ (as in Table \ref{tab:app}), then the false alarm
probability is
\begin{equation}\label{eq:PFA}
F=1-(1-{\rm Prob}(z>z_0))^M,
\end{equation}
where $M$ is the number of independent frequencies that were
examined. In Figure \ref{fig:dist}, we show the distribution of
maximum periodogram powers for sets of evenly spaced data with $N=20$.
The crosses are the results of our Monte Carlo simulations. We used
three different normalizations for the periodogram. The lines show the
theoretical distributions. In the case where the noise variance
$\sigma_0^2$ is known, we use equation (\ref{eq:PFA}) with ${\rm
Prob}(z>z_0)=\exp(-z_0)$ and $M=N$ (dotted line). For the
normalizations by $s^2$ and $s_n^2$ (dot-dashed and dashed lines), we
use the distributions given in Table \ref{tab:app} and find the best
fit value of $M$, fitting to the tail (${\rm Prob}>0.5$) of the
distribution. Both distributions give the same value, $M=23.5$. Notice
that normalizing by $s_n^2$ broadens the distribution of maximum
powers (because of the extra uncertainty in the value of $s_n^2$),
whereas normalizing by $s^2$ narrows the distribution (because of the
correlation with $s_f^2$).

Horne \& Baliunas (1986) used Monte Carlo simulations to find the
false alarm probability for evenly-spaced data sets\footnote{Note that
due to a typographical error, the values of $M$ given in the {\it
tables} in HB86 are incorrect (Baliunas 1998, private
communication).}. Normalizing by the sample variance, they assumed
${\rm Prob}(z>z_0)=\exp(-z_0)$ and fit for $M$ as a function of
$N$. However, as Figure \ref{fig:dist} shows, the distribution of $z$
is different from exponential, especially in the tails of the
distribution. {\it Thus the relations of Horne \& Baliunas (1986) give
inaccurate estimates of false alarm probabilities or detection
thresholds.} Because the distribution is squashed, the effect is to
{\it overestimate} both detection thresholds and false alarm
probabilities.

In agreement with Press et al.  (1992) and Marcy \& Benitz (1989),
Horne \& Baliunas found that $M\approx N$ when the period range
searched was from the average Nyquist period to the duration of the
data set. However, in this paper, we evaluate frequencies several
times larger than the average Nyquist frequency. A naive estimate of
the number of independent frequencies is $T\Delta f$. Our numerical
results show that in general the number of independent frequencies is
less than this estimate, hence our Monte Carlo approach for finding
the false alarm probabilities in \S \ref{sec:periodicity}. It would be
useful to have a method, for a given data set, of estimating the
number of independent frequencies, allowing one to write down false
alarm probabilities analytically. One approach may be to look at the
correlations between residuals (Schwarzenberg-Czerny 1991).

Finally, we discuss the difference between normalizing by $s_n^2$ at
each frequency, and normalizing by $s_n^2$ evaluated at the {\it best
fit} frequency. This choice is really a matter of taste. The
distribution of maximum periodogram powers is the same, by
definition. However, it seems to us that it is fairer to make
comparisons between frequencies using the same normalization for the
noise in $\chi^2$. Thus our choice of normalization is that of
Gilliland \& Baliunas (1987, eq. [\ref{eq:z}]): we normalize by the
same factor for each frequency, namely $s_n^2$ evaluated at the best
fit period.


\clearpage
\begin{deluxetable}{lcccccccl}
\footnotesize
\tablecaption{Orbits of Known Planetary-Mass Companions\label{tab:planets}}
\tablewidth{0pt}
\tablehead{
\colhead{Star} & \colhead{Lick} & \colhead{Spectral} &
\colhead{$P$} & \colhead{$a$} & \colhead{$K$} & \colhead{$M_p \sin i$} & \colhead{$e$} &
\colhead{Reference}\\
\colhead{Name} & \colhead{Survey?\tablenotemark{a}} & \colhead{Type} &
\colhead{(days)} & \colhead{(AU)} & \colhead{(m/s)} & \colhead{($M_J$)} & \colhead{} & \colhead{}
}
\startdata
HD 187123       & & G3V  & 3.10  & 0.042 & 83 & 0.57 & 0.03 & Butler et al. 1998\nl
$\tau$ Boo (HR5185) & Y & F7V   & 3.31  & 0.047 & 468 & 3.66 & 0.00 & Butler et al. 1997 \nl
51 Peg (HR8729) & & G5V   & 4.23  & 0.051 & 56 & 0.44 & 0.01 & Mayor \& Queloz 1995 \nl
&&&&&&&& Marcy et al. 1997 \nl
Ups And (HR458)  & Y & F8V   & 4.62  & 0.054 & 71.9 & 0.61 & 0.15 & Butler et al. 1997 \nl
HD 217107       & & G7V    & 7.12 & 0.072 & 140 & 1.28 & 0.14 & Fischer et al. 1999\nl
$\rho^1$ 55 Cnc (HR3522) & Y & G8V   & 14.7  & 0.11  & 75.9 & 0.85 & 0.04 & Butler et al. 1997 \nl
Gliese 86       & & K1V  & 15.8  & 0.114 & 379 & 4.9 & 0.04 & Queloz et al. 1999\nl
HD 195019       & & G3V/IV & 18.3 & 0.136 & 275 & 3.43 & 0.03 & Fischer et al. 1999\nl
$\rho$ CrB (HR5968) & & G2V   & 39.6  & 0.23  & 67 & 1.1 & 0.11 & Noyes et al. 1997 \nl
Gliese 876  & Y & M4    & 61    & 0.21  & 217 & 2.1 & 0.27 & Marcy et al. 1998 \nl
&&&&&&&& Delfosse et al. 1998 \nl
HD 168443       & & G8IV   & 57.9 & 0.28  & 350 & 4.96 & 0.54 & Marcy et al. 1999\nl
HD 114762       & & F7V & 84.0  & 0.41  & 618 & 11.0 & 0.33 & Latham et al. 1989\nl
&&&&&&&& Marcy et al. 1999 \nl
70 Vir (HR5072) &Y & G2.5V & 117 & 0.47  & 316 & 7.4 & 0.40 & Marcy \& Butler 1996 \nl
HD 210277       & & G7V    & 437  & 1.20  & 42  & 1.28 & 0.45 & Marcy et al. 1999\nl
16 Cyg B (HR7504) &Y & G2.5  & 799 & 1.6  & 50.3 & 1.67 & 0.69 & Cochran et al. 1997 \nl
47 UMa (HR4277) & Y & G0V   & 1092  & 2.1  & 47 & 2.38 & 0.11 & Butler \& Marcy 1996 \nl
14 Her          & & K0V   & 1620  & 2.5   & 75 & 3.3 & 0.36 & Mayor et al. 1998 \nl
\enddata
\tablenotetext{a}{We indicate with a ``Y'' those companions discovered by the original Lick survey.}
\end{deluxetable}


\begin{deluxetable}{llcrrrrr} 
\small
\tablecaption{Summary of the Observations\label{tab:stars}} 
\tablewidth{0pt}
\tablehead{
\multicolumn{2}{c}{Star} & \colhead{Spec.} &
\colhead{$P_{\rm rot}$\tablenotemark{a}} &
\colhead{$<\sigma_D>\tablenotemark{b}$} &
 \colhead{$\sigma_{\rm rms}$} &
 \colhead{$N$} &
\colhead{$T$} \nl
\colhead{HR} & \colhead{HD} & \colhead{Type} &
\colhead{(d)} & \colhead{(${\rm m\ s^{-1}}$)} &
\colhead{(${\rm m\ s^{-1}}$)} & \colhead{} & \colhead{(yr)} \nl
}
\startdata
8 & 166 &K0 & 5.7& 20/7.4 & 18.5 & 41 & 11.4 \nl
88 & 1835 &G3 & 8& 26/15 & 40.8 & 56 & 11.2 \nl
166 & 3651 &K0 & 44& 15/7.1 & 12.2 & 70 & 11.2 \nl
219a & 4614 &G0 & 15& 23/6.2 & 40.5 & 69 & 11.1 \nl
222 & 4628 &K2 & 39& 18/7.0 & 14.4 & 40 & 11.2 \nl
458 & 9826 &F8 & 12& 29/14 & 74.2 & 110 & 11.2 \nl
493 & 10476 &K1 & 35& 16/6.0 & 8.93 & 38 & 10.3 \nl
509 & 10700 &G8 & 34& 18/6.6 & 10.3 & 278 & 11.2 \nl
582 & 12235 &G2 & 14& 23/8.8 & 15.4 & 34 & 10.4 \nl
753 & 16160 &K3 & 48& 17/6.9 & 38.7 & 33 & 10.4 \nl
799 & 16895 &F8 & 7& 26/13 & 15.5 & 43 & 11.2 \nl
857 & 17925 &K2 & 6.9& 18/9.4 & 25.4 & 27 & 11.2 \nl
937 & 19373 &G0 & 21& 18/6.5 & 11.5 & 131 & 11.2 \nl
962 & 19994 &F8 & 10 & 24/14 & 30.1 & 29 & 11.2 \nl
996 & 20630 &G5 & 9& 22/7.5 & 24.4 & 67 & 11.2 \nl
1084 & 22049 & K2 & 12& 19/5.9 & 19.2 & 121 & 11.2 \nl
1101 & 22484 &F9 & 18& 19/6.8 & 15.2 & 40 & 11.2 \nl
1262 & 25680 &G5 & 9& 22/7.7 & 20.7 & 30 & 11.2 \nl
1325 & 26965 &K1 & 43& 22/6.6 & 22.4 & 66 & 11.2 \nl
1614 & 32147 &K3 & 47& 17/5.8 & 7.67 & 30 & 11.2 \nl
1729 & 34411 &G2 & 24& 17/7.8 & 9.43 & 106 & 11.2 \nl
1925 & 37394 &K1 & 11& 19/6.7 & 19.1 & 12 & 4.91 \nl
2047 & 39587 &G0 & 5.2& 43/13 & 1430 & 36 & 11.2 \nl
2483 & 48682 &G0 & 14& 22/11 & 13.6 & 61 & 11.2 \nl
2643 & 52711 &G4 & 19& 26/8.3 & 19.9 & 36 & 11.2 \nl
3262 & 69897 &F6 & 2.9 & 32/13 & 22.4 & 56 & 11.2 \nl
3522 & 75732 &G8 & 44& 16/7.9 & 60.7 & 114 & 8.91 \nl
3538 & 76151 &G3 & 15& 24/8.4 & 16 & 32 & 8.81 \nl
3625 & 78366 &F9 & 10& 26/11 & 24.2 & 28 & 9 \nl
3881 & 84737 &G0 & 23& 18/8.4 & 12.1 & 62 & 10.9 \nl
3951 & 86728 &G2 & 27& 17/7.9 & 11.4 & 75 & 10.5 \nl
4112 & 90839 &F8 & 7.3& 32/7.2 & 19.7 & 35 & 8.28 \nl
4277 & 95128 &G0 & 16& 22/7.5 & 31.1 & 86 & 11.4 \nl
4345 & 97334 &G0 & 8& 25/16 & 30.4 & 36 & 10.5 \nl
4496 & 101501 &G8 & 17& 21/8.7 & 14.9 & 72 & 11 \nl
4540 & 102870 &F9 & 14& 25/6.8 & 17.5 & 80 & 10.5 \nl
4983 & 114710 &G0 & 12& 24/7.0 & 27 & 100 & 9.71 \nl
5011 & 115383 &G0 & 3.5& 29/13 & 26.5 & 40 & 8.89 \nl
5019 & 115617 &G6 & 29& 20/6.9 & 15.4 & 46 & 7.21 \nl
5072 & 117176 &G2 & 36& 19/7.4 & 185 & 92 & 10.4 \nl
5185 & 120136 &F7 & 4& 61/24 & 331 & 80 & 11.1 \nl
5273 & 122742 &G8 & 30& 25/8.6 & 4300 & 24 & 9.71 \nl
5384 & 126053 &G1 & 22& 24/9.0 & 15.1 & 29 & 10.1 \nl
5447 & 128167 &F2 & 0.3 & 53/24 & 51.8 & 48 & 10.3 \nl
5534 & 130948 &G0 & 3.2 & 29/14 & 33.4 & 24 & 10.1 \nl
5544a & 131156 &G8 & 6 & 23/8.1 & 67 & 60 & 11.0 \nl
5544b & 131156b &K4 & 11& 21/13 & 91.3 & 23 & 10.1 \nl
5553 & 131511 &K2 & 10 & 18/11 & 8880 & 20 & 7.85 \nl
5568 & 131977 &K4 & 40 & 18/9.7 & 11.3 & 47 & 5.06 \nl
5868 & 141004 &G0 & 26& 18/7.0 & 10.8 & 106 & 10.9 \nl
5914 & 142373 &F8 & 15& 26/8.8 & 15.9 & 83 & 10.8 \nl
5933 & 142860 &F6 & 3 & 56/19 & 64.9 & 70 & 9.15 \nl
5968 & 143761 &G2 & 30& \nodata/8.5 & 31 & 20 & 1.21 \nl
6171 & 149661 &K2 & 21& 17/7.8 & 11.2 & 40 & 7.31 \nl
6458 & 157214 &G0 & 22& 21/9.6 & 14.3 & 72 & 11.2 \nl
6623 & 161797 &G5 & 42& 13/5.4 & 146 & 76 & 11 \nl
6806 & 166620 &K2 & 43& 17/7.4 & 14.9 & 39 & 11.1 \nl
6869 & 168723 &K2 & 44 & 14/5.4 & 15 & 47 & 11 \nl
7061 & 173667 &F6 & 2.3 & 64/36 & 129 & 78 & 11.2 \nl
7462 & 185144 &K0 & 27& 16/6.4 & 12.2 & 42 & 9.3 \nl
7503 & 186408 &G1 & 27& 19/10 & 18.8 & 62 & 11.4 \nl
7504 & 186427 &G2 & 29& 21/11 & 29.4 & 124 & 11.4 \nl
7602 & 188512 &G8 & 52& 15/5.7 & 13.2 & 85 & 11.4 \nl
7672 & 190406 &G1 & 14& 21/9.7 & 87.7 & 74 & 11 \nl
8085 & 201091 &K5 & 35& 17/6.0 & 10.6 & 100 & 11.4 \nl
8086 & 201092 &K7 & 38& 18/7.2 & 31.9 & 58 & 11.4 \nl
8314 & 206860 &G0 & 5& 37/19 & 33.8 & 36 & 9.89 \nl
8382 & 208801 &K2 & 45 & 17/6.1 & 7.77 & 21 & 5.29 \nl
8665 & 215648 &F6 & 3.6 & 30/16 & 26.9 & 61 & 11 \nl
8729 & 217014 & G5 & 22& \nodata/11 & 40.5 & 220 & 2.77 \nl
8832 & 219134 &K3 & 48& 16/7.6 & 13.4 & 45 & 5.86 \nl
8969 & 222368 &F7 & 9& 26/10 & 23.7 & 60 & 9.89 \nl
GL250a & 50281 &K3 & 45 & 17/7.6 & 15.1 & 18 & 5.01 \nl
GL641 & 152391 &G6 & 11& 13/11 & 24 & 29 & 5.18 \nl
GL688 & 160346 &K3 & 37& 11/7.1 & 3730 & 7 & 3.99 \nl
GL716 & 170657 &K3 & 15 & 8.6/16 & 15.8 & 13 & 4.87 \nl

\enddata \tablenotetext{a}{Rotation periods are from Baliunas,
Sokoloff \& Soon 1996, Soderblom 1985, or Fischer 1999 (private
communication).}  \tablenotetext{b}{We give the mean internal Doppler
error before and after November 1994. The internal errors have been
augmented as described in \S\ref{sec:errors}.}
\end{deluxetable}


\begin{deluxetable}{rlrrrr} 
\small
\tablecaption{Test for Long Term Trends ($F$-Test)\label{tab:trends}\tablenotemark{a}}
\tablewidth{0pt}
\tablehead{
\colhead{} & \colhead{Star} & \colhead{Slope} &
\colhead{$N$} & \colhead{$F_{1,N-2}$} & \colhead{False}\nl
\colhead{} & \colhead{(HR)} & \colhead{(${\rm m\ s^{-1}\ yr^{-1}}$)} & 
\colhead{} & \colhead{} & \colhead{Alarm} 
}
\startdata
&8 & -1.5$\pm$1.7 &41 & 1.8 & 0.187\nl
&88 & -1.1$\pm$1.4 &56 & 0.206 & 0.652\nl
&166 & 0.68$\pm$0.58 &70 & 0.835 & 0.364\nl
$\ast$ &219a & 9.3$\pm$0.58 &69 & 681 & 8.13e-37\nl
&222 & 0.76$\pm$0.72 &40 & 1.31 & 0.259\nl
&458 & -6.6$\pm$0.82 &110 & 3.47 & 0.0651\nl
&493 & 1.4$\pm$0.68 &38 & 9.36 & 0.00418\nl
&509 & 0.82$\pm$0.29 &278 & 16.3 & 7.1e-05\nl
&582 & -3.9$\pm$1.3 &34 & 28.5 & 7.33e-06\nl
$\ast$ &753 & 12$\pm$0.73 &33 & 399 & 2.95e-19\nl
&799 & -1.6$\pm$1.4 &43 & 4.77 & 0.0348\nl
&857 & 3.5$\pm$2.2 &27 & 2.09 & 0.161\nl
&937 & 1.2$\pm$0.39 &131 & 20.4 & 1.4e-05\nl
&962 & 3.3$\pm$1.2 &29 & 4.35 & 0.0466\nl
&996 & -3.1$\pm$0.89 &67 & 9.56 & 0.0029\nl
&1084 & 1.4$\pm$0.44 &121 & 10.4 & 0.00161\nl
&1101 & 0.2$\pm$0.71 &40 & 0.118 & 0.734\nl
&1262 & 0.015$\pm$1.4 &30 & 0.000103 & 0.992\nl
$\ast$ &1325 & 5$\pm$0.45 &66 & 152 & 1.49e-18\nl
&1614 & 1.7$\pm$0.89 &30 & 6.42 & 0.0172\nl
&1729 & 1.6$\pm$0.39 &106 & 29.8 & 3.3e-07\nl
&1925 & -4.6$\pm$2.8 &12 & 2.38 & 0.154\nl
$\ast$ &2047 & 310$\pm$1.8 &36 & 232 & 9.5e-17\nl
&2483 & -0.31$\pm$0.77 &61 & 0.317 & 0.576\nl
&2643 & 2.2$\pm$1.2 &36 & 5.12 & 0.0301\nl
&3262 & 1.9$\pm$1.1 &56 & 3.83 & 0.0557\nl
&3522 & -3.7$\pm$0.67 &114 & 0.442 & 0.508\nl
&3538 & 1.2$\pm$1.5 &32 & 0.991 & 0.327\nl
&3625 & 0.79$\pm$1.9 &28 & 0.153 & 0.699\nl
&3881 & 0.2$\pm$0.57 &62 & 0.205 & 0.652\nl
&3951 & 1.8$\pm$0.63 &75 & 10.4 & 0.00185\nl
&4112 & -2.9$\pm$1.5 &35 & 8.75 & 0.00569\nl
&4277 & 2.2$\pm$0.69 &86 & 2.19 & 0.142\nl
&4345 & 1.7$\pm$1.9 &36 & 0.447 & 0.508\nl
&4496 & 1.3$\pm$0.58 &72 & 6.69 & 0.0118\nl
&4540 & 1.3$\pm$0.62 &80 & 3.93 & 0.051\nl
&4983 & 3.2$\pm$0.68 &100 & 20.3 & 1.83e-05\nl
&5011 & -1.8$\pm$3.3 &40 & 0.893 & 0.351\nl
&5019 & 4.4$\pm$0.88 &46 & 43.8 & 4.13e-08\nl
&5072 & -3.1$\pm$0.55 &92 & 0.073 & 0.788\nl
&5185 & -23$\pm$1.8 &80 & 3.43 & 0.0677\nl
&5273 & 850$\pm$1.5 &24 & 13.5 & 0.00132\nl
&5384 & -1.3$\pm$1.1 &29 & 3.09 & 0.0902\nl
&5447 & -2.2$\pm$1.8 &48 & 1.21 & 0.277\nl
&5534 & -3.5$\pm$1.7 &24 & 1.61 & 0.217\nl
$\ast$ &5544a & 18$\pm$1.1 &60 & 197 & 2.8e-20\nl
$\ast$ &5544b & -25$\pm$1.2 &23 & 187 & 6.27e-12\nl
&5553 & 1000$\pm$1.3 &20 & 2.46 & 0.134\nl
&5568 & -3.6$\pm$1.3 &47 & 7.32 & 0.00959\nl
&5868 & -0.98$\pm$0.49 &106 & 4.27 & 0.0413\nl
&5914 & 1.5$\pm$0.64 &83 & 13.5 & 0.000422\nl
&5933 & -8.1$\pm$1.6 &70 & 20.2 & 2.78e-05\nl
&5968 & -19$\pm$5.5 &20 & 0.957 & 0.341\nl
&6171 & -0.56$\pm$1.1 &40 & 0.586 & 0.449\nl
&6458 & 1.8$\pm$0.58 &72 & 20.2 & 2.71e-05\nl
$\ast$ &6623 & -47$\pm$0.41 &76 & 11800 & 2.05e-83\nl
&6806 & 1.3$\pm$0.68 &39 & 3.3 & 0.0775\nl
&6869 & 0.19$\pm$0.69 &47 & 0.0464 & 0.83\nl
&7061 & -8.2$\pm$2 &78 & 3.07 & 0.0836\nl
&7462 & 0.86$\pm$0.6 &42 & 2.63 & 0.113\nl
&7503 & -3$\pm$0.7 &62 & 18.2 & 7.05e-05\nl
&7504 & -1.9$\pm$0.69 &124 & 1.29 & 0.258\nl
&7602 & 2$\pm$0.42 &85 & 25 & 3.21e-06\nl
$\ast$ &7672 & -24$\pm$0.6 &74 & 2310 & 1.92e-56\nl
&8085 & 0.59$\pm$0.4 &100 & 4.5 & 0.0364\nl
$\ast$ &8086 & 8.9$\pm$0.59 &58 & 326 & 4.96e-25\nl
&8314 & -6.8$\pm$2.9 &36 & 9.95 & 0.00335\nl
&8382 & 1.5$\pm$1.4 &21 & 1.99 & 0.174\nl
&8665 & -3.7$\pm$1.2 &61 & 11.7 & 0.00112\nl
&8729 & -2.3$\pm$1.8 &220 & 0.106 & 0.745\nl
&8832 & 2.8$\pm$1.1 &45 & 9.02 & 0.00444\nl
&8969 & -1$\pm$1.1 &60 & 1.2 & 0.277\nl
&GL250a & -0.53$\pm$1.5 &18 & 0.0803 & 0.781\nl
&GL641 & 0.35$\pm$2.1 &29 & 0.00937 & 0.924\nl
&GL688 & -2200$\pm$2.5 &7 & 6.5 & 0.0514\nl
&GL716 & 0.38$\pm$2.2 &13 & 0.0353 & 0.854\nl
\enddata 
\tablenotetext{a}{We mark with ``$\ast$'' in the leftmost column those
stars for which we subtract the slope from the data (\S
\ref{sec:trends}). These stars have a false alarm probability
$<10^{-5}$ and a slope greater than $5\msyr$.}
\end{deluxetable}


\begin{deluxetable}{rlrrrr} 
\small
\tablecaption{Test for Excess Variability ($\chi^2$-Test)\label{tab:variability}\tablenotemark{a}}
\tablewidth{0pt}
\tablehead{
\colhead{} & \colhead{Star} & \colhead{Expected} & \colhead{Observed} &
\colhead{$\chi^2_\nu$} & \colhead{False} \nl
\colhead{} & \colhead{(HR)} & \colhead{Rms\tablenotemark{b}} & 
\colhead{Rms\tablenotemark{c}} &
\colhead{} & \colhead{Alarm} \nl
\colhead{} & \colhead{} & \colhead{(${\rm m\ s^{-1}}$)} & 
\colhead{(${\rm m\ s^{-1}}$)} &
\colhead{} & \colhead{}
}
\startdata
& 8 & 28.5 & 18.5 & 0.423 & 0.999\nl
$\ast$ & 88 & 25.3 & 40.8 & 2.78 & 3.94e-11\nl
$\ast$ & 166 & 10.2 & 12.2 & 1.64 & 0.000676\nl
& 219a & 21.1 & 18.6 & 0.379 & 1\nl
& 222 & 14 & 14.4 & 0.853 & 0.728\nl
$\ast$c & 458 & 21 & 74.2 & 19.3 & 0\nl
& 493 & 12.6 & 8.93 & 0.551 & 0.988\nl
& 509 & 14.2 & 10.3 & 0.512 & 1\nl
& 582 & 19.3 & 15.4 & 0.591 & 0.97\nl
& 753 & 13.3 & 10.7 & 0.622 & 0.95\nl
& 799 & 26.7 & 15.5 & 0.283 & 1\nl
& 857 & 24 & 25.4 & 1.33 & 0.118\nl
& 937 & 16.8 & 11.5 & 0.571 & 1\nl
& 962 & 23.9 & 30.1 & 1.59 & 0.024 \nl
& 996 & 23.2 & 24.4 & 1.43 & 0.0122\nl
& 1084 & 19.9 & 19.2 & 1.07 & 0.29\nl
& 1101 & 17.6 & 15.2 & 0.662 & 0.948\nl
& 1262 & 20.9 & 20.7 & 0.976 & 0.502\nl
& 1325 & 17.9 & 19 & 0.795 & 0.883\nl
& 1614 & 11.1 & 7.67 & 0.645 & 0.929\nl
& 1729 & 13 & 9.43 & 0.712 & 0.989\nl
& 1925 & 17.8 & 19.1 & 1.3 & 0.217\nl
$\ast$ & 2047 & 42.7 & 364 & 125 & 0\nl
& 2483 & 19.1 & 13.6 & 0.504 & 1\nl
& 2643 & 20.1 & 19.9 & 0.738 & 0.87\nl
& 3262 & 56.5 & 22.4 & 0.14 & 1 \nl
$\ast$c & 3522 & 9.81 & 60.7 & 70.1 & 0\nl
& 3538 & 19.2 & 16 & 0.716 & 0.877\nl
& 3625 & 22.4 & 24.2 & 1.07 & 0.369\nl
& 3881 & 15.7 & 12.1 & 0.585 & 0.996\nl
& 3951 & 12.4 & 11.4 & 0.868 & 0.784\nl
& 4112 & 31.6 & 19.7 & 0.528 & 0.989\nl
$\ast$c & 4277 & 16.1 & 31.1 & 4.83 & 4.32e-44\nl
$\ast$ & 4345 & 25.4 & 30.4 & 1.78 & 0.00302\nl
& 4496 & 17.4 & 14.9 & 0.777 & 0.918\nl
& 4540 & 21.9 & 17.5 & 1.08 & 0.303\nl
& 4983 & 23.9 & 27 & 1.3 & 0.0242\nl
& 5011 & 46.6 & 26.5 & 0.315 & 1\nl
& 5019 & 16.8 & 15.4 & 1.12 & 0.27\nl
$\ast$c & 5072 & 13.2 & 185 & 432 & 0\nl
$\ast$c & 5185 & 56.4 & 331 & 48.9 & 0\nl
$\ast$ & 5273 & 22.6 & 4300 & 34700 & 0\nl
& 5384 & 20.4 & 15.1 & 0.452 & 0.994\nl
& 5447 & 955 & 51.8 & 0.003 & 1 \nl
& 5534 & 49.5 & 33.4 & 0.44 & 0.99 \nl
&5544a & 30.1 & 34.1 & 1.3 & 0.06 \nl
$\ast$ & 5544b & 21.2 & 33.5 & 2.15 & 0.0017\nl
$\ast$ & 5553 & 20 & 8880 & 1.75e5 & 0 \nl
& 5568 & 11.2 & 11.3 & 1.06 & 0.371 \nl
& 5868 & 12.3 & 10.8 & 0.954 & 0.615\nl
& 5914 & 23.9 & 15.9 & 0.456 & 1\nl
& 5933 & 67.7 & 64.9 & 0.665 & 0.986 \nl
$\ast$c & 5968 & 9.23 & 31 & 12.3 & 4.62e-39\nl
& 6171 & 14.5 & 11.2 & 0.479 & 0.998\nl
& 6458 & 17.7 & 14.3 & 0.602 & 0.997\nl
& 6623 & 11.5 & 13.2 & 1.15 & 0.183\nl
& 6806 & 14.3 & 14.9 & 1.22 & 0.168\nl
& 6869 & 11.7 & 15 & 1.46 & 0.024 \nl
$\ast$& 7061 & 88.5 & 129 & 1.99 & 5e-7 \nl
& 7462 & 13.3 & 12.2 & 0.794 & 0.824\nl
& 7503 & 14.1 & 18.8 & 1.28 & 0.0674\nl
$\ast$c & 7504 & 14.2 & 29.4 & 5.82 & 2.32e-84\nl
& 7602 & 12.1 & 13.2 & 1.16 & 0.149\nl
& 7672 & 19.5 & 15.9 & 0.728 & 0.96\nl
& 8085 & 15.3 & 10.6 & 0.505 & 1\nl
& 8086 & 15 & 11.1 & 0.694 & 0.961\nl
& 8314 & 42.1 & 33.8 & 0.705 & 0.904\nl
& 8382 & 12 & 7.77 & 0.565 & 0.938 \nl
& 8665 & 47.1 & 26.9 & 0.29 & 1 \nl
$\ast$c & 8729 & 12.5 & 40.5 & 14.8 & 0\nl
& 8832 & 12.4 & 13.4 & 0.931 & 0.602\nl
& 8969 & 25.8 & 23.7 & 0.674 & 0.974\nl
& GL250a & 12.3 & 15.1 & 1.33 & 0.163 \nl
$\ast$ & GL641 & 15.7 & 24 & 2.8 & 1.14e-06\nl
$\ast$ & GL688 & 9.4 & 3730 & 21800 & 0\nl
& GL716 & 16.3 & 15.8 & 0.545 & 0.886 \nl

\enddata
\tablenotetext{a}{
We give the reduced $\chi^2$ either about the mean
or from the straight line fit (see \S \ref{sec:variability}).
We mark with ``$\ast$'' in the leftmost column those stars which show
a false alarm probability $<1\%$, and with a ``c'' those with
confirmed planetary-mass companions (Table \ref{tab:planets}).}
\tablenotetext{b}{The predicted rms taking into account the Doppler
errors and intrinsic variability.}
\tablenotetext{c}{The observed rms of the velocities after subtraction of the
mean or straight line.}
\end{deluxetable}


\begin{deluxetable}{rlrrrrrr} 
\small
\tablecaption{Periodogram Results\label{tab:periodogram}\tablenotemark{a}}
\tablewidth{0pt}
\tablehead{
\colhead{} & \colhead{Star} & \colhead{$z_{\rm max}$} &
\multicolumn{3}{c}{False Alarm Probability} & \colhead{Period}&
\colhead{$K$} \nl
\colhead{} & \colhead{(HR)} & \colhead{} & \colhead{Gaussian} &
\colhead{Mixed} & \colhead{Analytic\tablenotemark{b}} 
& \colhead{(days)}& \colhead{(m/s)}
}
\startdata
&8 & 14.2& 0.1& 0.1& 0.0539& 6.31& 15.1\nl
&88 & 15& 0.03& 0.03& 0.0154& 3.1& 31.8\nl
&166 & 13.9& 0.04& 0.03& 0.0135& 4.01& 7.25\nl
&219a & 11& 0.33& 0.76& 0.19& 6.4& 6.72\nl
&222 & 9.93& 0.34& 0.28& 0.288& 7.9& 8.43\nl
$\ast$c &458 & 38.2& $<0.0025$& $<0.0025$& 6.2e-10& 4.62& 64.4\nl
&493 & 16.2& 0.03& 0.1& 0.00993& 7.89& 6.62\nl
$\ast$ &509 & 35.4& $<0.0025$& $<0.0025$& 5.95e-11& 19.5& 4.1\nl
&582 & 12.4& 0.28& 0.2& 0.174& 11000& 22.1\nl
&753 & 8.32& 0.63& 0.63& 0.77& 2.27& 7.06\nl
&799 & 15.9& 0.06& 0.06& 0.0366& 15.5& 11.7\nl
&857 & 23.5& 0.0175& 0.03& 0.00253& 8.26& 36.7\nl
$\ast$ &937 & 18& $<0.0025$& 0.0125& 0.000479& 1920& 5.73\nl
&962 & 16.3& 0.08& 0.28& 0.044& 3.28& 28.6\nl
$\ast$ &996 & 40.9& $<0.0025$& $<0.0025$& 4.71e-09& 46.9& 28.7\nl
$\ast$ &1084 & 36.1& $<0.0025$& $<0.0025$& 1.35e-09& 2520& 14.7\nl
&1101 & 10.3& 0.42& 0.49& 0.36& 3.73& 8.35\nl
&1262 & 12.2& 0.4& 0.34& 0.398& 19& 19.2\nl
&1325 & 16.1& 0.015& 0.207& 0.00232& 5.39& 8.2\nl
&1614 & 9.93& 0.34& 0.67& 0.283& 92.6& 5.2\nl
$\ast$ &1729 & 59.1& $<0.0025$& $<0.0025$& 0& 376& 8.12\nl
&1925 & 45.1& 0.0175& 0.12& 0.000584& 3.2& 85.7\nl
$\ast$ &2047 & 498& $<0.0025$& $<0.0025$& 0& 11000& 3410\nl
&2483 & 12.6& 0.11& 0.2& 0.0585& 6.89& 10.5\nl
&2643 & 9.88& 0.36& 0.2& 0.35& 3640& 14.4\nl
&3262 & 13.2& 0.09& 0.15& 0.0619& 2.54& 16.4\nl
$\ast$c &3522 & 978& $<0.0025$& $<0.0025$& 0& 14.6& 75.9\nl
&3538 & 9.43& 0.58& 0.62& 0.625& 4.89& 10.3\nl
&3625 & 18.4& 0.01& 0.05& 0.0083& 29.1& 24.1\nl
&3881 & 13& 0.04& 0.02& 0.0347& 15.6& 7.57\nl
$\ast$&3951 & 15.6& 0.005& 0.0625& 0.00323& 6.4& 7.57\nl
&4112 & 13.2& 0.15& 0.13& 0.0633& 8.84& 17.5\nl
$\ast$c &4277 & 456& $<0.0025$& $<0.0025$& 0& 1070& 46.4\nl
$\ast$ &4345 & 26.1& $<0.0025$& 0.0025& 0.000289& 2.96& 33.5\nl
$\ast$ &4496 & 21.6& $<0.0025$& $<0.0025$& 4.95e-05& 2.03& 11.7\nl
$\ast$ &4540 & 27.1& $<0.0025$& $<0.0025$& 8.9e-07& 10.7& 14.8\nl
$\ast$ &4983 & 27.7& $<0.0025$& $<0.0025$& 3.87e-07& 352& 24.8\nl
&5011 & 14.3& 0.09& 0.04& 0.0298& 2.23& 25.4\nl
$\ast$ &5019 & 29.6& $<0.0025$& 0.11& 2.88e-06& 3550& 14.6\nl
$\ast$c &5072 & 304& $<0.0025$& $<0.0025$& 0& 117& 252\nl
$\ast$c &5185 & 387& $<0.0025$& $<0.0025$& 0& 3.31& 456\nl
$\ast$ &5273 & 62.4& $<0.0025$& $<0.0025$& 1.24e-06& 3300& 4130\nl
&5384 & 10.4& 0.35& 0.45& 0.336& 14.9& 8.83\nl
&5447 & 19.3& 0.0125& 0.0075& 0.00222& 2.08& 50.2\nl
&5534 & 11.3& 0.71& 0.79& 0.805& 2.88& 35.1\nl
$\ast$&5544a & 20.1& 0.0025& 0.0375& 0.000672& 3.78& 26.6\nl
&5544b & 14.4& 0.38& 0.57& 0.313& 9.9& 29\nl
&5553 & 28.7& 0.02& 0.177& 0.00546& 126& 10500\nl
$\ast$ &5568 & 18.6& $<0.0025$& $<0.0025$& 0.00091& 4.06& 9.67\nl
&5868 & 13.9& 0.01& 0.07& 0.0123& 358& 6.06\nl
&5914 & 14& 0.02& 0.2& 0.0107& 39.6& 7.52\nl
&5933 & 15.7& 0.01& 0.223& 0.00498& 9.04& 45.9\nl
$\ast$c &5968 & 129& $<0.0025$& $<0.0025$& 1.89e-08& 40& 65.7\nl
&6171 & 12.7& 0.07& 0.05& 0.0566& 2.14& 10.3\nl
&6458 & 13& 0.04& 0.34& 0.0271& 11000& 16.1\nl
$\ast$ &6623 & 36.2& $<0.0025$& $<0.0025$& 2.07e-08& 11000& 55\nl
&6806 & 12.8& 0.11& 0.15& 0.0561& 32& 8.83\nl
&6869 & 13& 0.08& 0.28& 0.0314& 4.81& 8.03\nl
$\ast$&7061 & 17.4& 0.005& 0.0025& 0.00173& 5.32& 94.4\nl
&7462 & 10.2& 0.25& 0.16& 0.178& 16.9& 7.42\nl
&7503 & 11.6& 0.19& 0.19& 0.141& 11000& 24.9\nl
$\ast$c &7504 & 101& $<0.0025$& $<0.0025$& 0& 770& 29.3\nl
$\ast$ &7602 & 45.3& $<0.0025$& 0.08& 4.59e-11& 4730& 12.7\nl
$\ast$ &7672 & 18.9& $<0.0025$& 0.107& 0.000611& 3.19& 15\nl
&8085 & 13.9& 0.02& 0.1& 0.00822& 2.22& 4.55\nl
$\ast$ &8086 & 20.3& $<0.0025$& 0.03& 0.000368& 19& 8.58\nl
&8314 & 17.6& 0.03& 0.12& 0.0131& 2.68& 35.8\nl
&8382 & 9.79& 0.46& 0.66& 0.486& 6.19& 5.81\nl
&8665 & 12.1& 0.16& 0.28& 0.111& 42.8& 22.4\nl
$\ast$c &8729 & 4440& $<0.0025$& $<0.0025$& 0& 4.23& 56.2\nl
&8832 & 16.7& 0.0175& 0.307& 0.00601& 11000& 101\nl
&8969 & 8.36& 0.57& 0.26& 0.61& 4& 11.7\nl
&GL250a & 23.8& 0.05& 0.06& 0.0192& 20.7& 15.1\nl
&GL641 & 18.9& 0.0125& 0.02& 0.0116& 6.33& 33.7\nl
&GL688 & 269& 0.16& 0.21& 0.104& 2.36& 6370\nl
&GL716 & 7.21& 0.65& 0.92& 0.904& 3.27& 19.6\nl

\enddata \tablenotetext{a}{A ``$\ast$'' in the leftmost column
indicates those stars with a periodicity significant at the 1\%
level. A ``c'' indicates those stars with confirmed planetary-mass
companions (Table \ref{tab:planets}).}  \tablenotetext{b}{Determined
by fitting for the number of independent frequencies using the
analytic distribution given in Table \ref{tab:app}. A ``0'' means
infinitesimal false alarm probability.}
\end{deluxetable}


\begin{deluxetable}{rlrrr|rlrrr}
\small 
\tablecaption{Average Periodogram Power\tablenotemark{a}\label{tab:zav}}
\tablewidth{0pt}
\tablehead{
\colhead{} & \colhead{Star} & \colhead{$\bar{z}$} & 
\multicolumn{2}{c}{False Alarm}\vline &
\colhead{} & \colhead{Star} & \colhead{$\bar{z}$} & 
\multicolumn{2}{c}{False Alarm}\nl
\colhead{} & \colhead{(HR)} & \colhead{} &
\colhead{Gaussian} & \colhead{Mixed}\vline &
\colhead{} & \colhead{(HR)} & \colhead{} &
\colhead{Gaussian} & \colhead{Mixed}}
\startdata
&8 & 2.12& 0.05& 0.07 & $\ast$ &5019 & 4.92& $<0.0025$& 0.02\nl
$\ast$ &88 & 2.62& $<0.0025$& $<0.0025$ & $\ast$c &5072 & 19.2& $<0.0025$& $<0.0025$\nl
$\ast$ &166 & 1.98& $<0.0025$& 0.01 & $\ast$c &5185 & 22.6& $<0.0025$& $<0.0025$\nl
&219a & 1.4& 0.39& 0.71 & $\ast$ &5273 & 9.92& $<0.0025$& 0.02\nl
&222 & 1.63& 0.26& 0.12 & &5384 & 2.35& 0.15& 0.2\nl
$\ast$c &458 & 3.66& $<0.0025$& $<0.0025$ & &5447 & 3.12& 0.0025& 0.0025\nl
&493 & 2.33& 0.09& 0.11 & &5534 & 2& 0.55& 0.63\nl
$\ast$ &509 & 2.75& $<0.0025$& $<0.0025$ & $\ast$ &5544a & 3.29& $<0.0025$& $<0.0025$\nl
&582 & 1.62& 0.43& 0.39 & &5544b & 3.64& 0.12& 0.24\nl
&753 & 1.32& 0.63& 0.42 & &5553 & 4.35& 0.04& 0.198\nl
&799 & 2.26& 0.04& 0.02 & $\ast$ &5568 & 2.7& $<0.0025$& 0.0025\nl
&857 & 4.71& 0.0125& 0.0075 & $\ast$ &5868 & 1.84& $<0.0025$& 0.03\nl
$\ast$ &937 & 1.95& $<0.0025$& $<0.0025$ & $\ast$ &5914 & 1.94& $<0.0025$& 0.07\nl
&962 & 2.13& 0.19& 0.42 & $\ast$ &5933 & 2.4& $<0.0025$& 0.04\nl
$\ast$ &996 & 8.04& $<0.0025$& $<0.0025$ & $\ast$c &5968 & 19.9& $<0.0025$& $<0.0025$\nl
$\ast$ &1084 & 3.37& $<0.0025$& $<0.0025$ & &6171 & 1.7& 0.27& 0.15\nl
&1101 & 1.8& 0.33& 0.39 & &6458 & 1.49& 0.18& 0.35\nl
&1262 & 1.75& 0.53& 0.32 & $\ast$ &6623 & 2.99& $<0.0025$& 0.04\nl
&1325 & 2.7& 0.0175& 0.237 & &6806 & 2.94& 0.02& 0.02\nl
&1614 & 2.03& 0.14& 0.54 & &6869 & 2.69& 0.02& 0.1\nl
$\ast$ &1729 & 4.47& $<0.0025$& $<0.0025$ & $\ast$ &7061 & 2.56& $<0.0025$& $<0.0025$\nl
&1925 & 16.5& 0.005& 0.125 & &7462 & 1.69& 0.26& 0.09\nl
$\ast$ &2047 & 54.6& $<0.0025$& $<0.0025$ & &7503 & 1.56& 0.22& 0.12\nl
&2483 & 1.86& 0.05& 0.11 & $\ast$c &7504 & 6.6& $<0.0025$& $<0.0025$\nl
&2643 & 1.15& 0.72& 0.3 & $\ast$ &7602 & 7.05& $<0.0025$& $<0.0025$\nl
&3262 & 1.74& 0.07& 0.15 & &7672 & 2.5& 0.0025& 0.0075\nl
$\ast$c &3522 & 51.6& $<0.0025$& $<0.0025$ & &8085 & 2.26& 0.01& 0.02\nl
&3538 & 1.64& 0.53& 0.63 & &8086 & 3.09& 0.0025& 0.005\nl
&3625 & 3.07& 0.01& 0.04 & &8314 & 2.8& 0.01& 0.14\nl
&3881 & 1.44& 0.2& 0.12 & &8382 & 2.31& 0.27& 0.51\nl
&3951 & 2.51& 0.0025& 0.0025 & &8665 & 1.74& 0.14& 0.26\nl
&4112 & 2.46& 0.11& 0.04 & $\ast$c &8729 & 248& $<0.0025$& $<0.0025$\nl
$\ast$c &4277 & 19.1& $<0.0025$& $<0.0025$ & &8832 & 2.15& 0.05& 0.0675\nl
$\ast$ &4345 & 4.54& $<0.0025$& $<0.0025$ & &8969 & 1.22& 0.51& 0.11\nl
$\ast$ &4496 & 2.59& $<0.0025$& $<0.0025$ & &GL250a & 3.81& 0.14& 0.05\nl
$\ast$ &4540 & 5.31& $<0.0025$& $<0.0025$ & &GL641 & 3.04& 0.02& 0.01\nl
$\ast$ &4983 & 4.13& $<0.0025$& $<0.0025$ & &GL688 & 104& 0.18& 0.22\nl
&5011 & 2.43& 0.03& 0.02 & &GL716 & 1.44& 0.76& 0.96\nl

\enddata \tablenotetext{a}{Here we give the mean periodogram power
$\bar{z}$ evaluated by summing all the powers evaluated in a
periodogram and dividing by the number of frequencies. We mark with a
``$\ast$'' those stars with false alarm probability $<1$\%. A ``c''
indicates those stars with confirmed planetary-mass companions (Table
\ref{tab:planets}).}
\end{deluxetable}


\begin{deluxetable}{lcccccccc}
\small
\tablecaption{Stars with Significant Variability or Periodicities\tablenotemark{a}\label{tab:summary}}

\tablewidth{0pt}
\tablehead{
\colhead{Star} & \colhead{Slope\tablenotemark{b}} & \multicolumn{2}{c}{Variability\tablenotemark{c}} & 
\multicolumn{4}{c}{Periodogram\tablenotemark{d}} & \colhead{$P_{rot}$}\nl
\colhead{} & \colhead{(m/s/yr)} & \colhead{$\chi^2$} & \colhead{$\bar{z}$} &
\colhead{Period} & \colhead{$K$ (m/s)} & \colhead{$\sigma_{\rm exp}$ (m/s)} &
\colhead{$\sigma_{\rm rms}$ (m/s)} & \colhead{(days)}}
\startdata
\multicolumn{9}{c}{Stars with Significant Slopes}\nl
\nl
HR 219a & $9.3\pm 0.6$ & \nodata & \nodata & \nodata & \nodata & \nodata & \nodata &15 \nl
HR 753 & $12\pm 0.7$ & \nodata & \nodata & \nodata & \nodata & \nodata & \nodata & 48 \nl
HR 1325 & $5\pm 0.5$ & \nodata & \nodata & \nodata & \nodata & \nodata & \nodata & 43 \nl
HR 5544a & $18\pm 1.1$ & \nodata & $\ast$ & 3.8 d & 27 & 30 & 26 & 6 \nl
HR 5544b & $-25\pm 1.2$ & $\ast$ & \nodata & \nodata & \nodata & \nodata & \nodata &11 \nl
HR 7672 & $-24\pm 0.6$ & \nodata & \nodata & 3.2 d & 15 & 20 & 11 & 14 \nl
HR 8086 & $8.9\pm 0.6$ & \nodata & \nodata & 19 d & 8.6 & 15 & 6.7 &38 \nl
\nl
\multicolumn{9}{c}{Chromospherically Inactive Stars ($P_{rot}>14$ days)}\nl
\nl
HR 166  & \nodata & $\ast$ & $\ast$ & \nodata & \nodata & \nodata & \nodata &44 \nl
HR 509  & \nodata & \nodata & $\ast$ & 20 d & 4.1 & 14 & 6.0 &34 \nl
HR 937  & \nodata & \nodata & $\ast$ & 5.3 yr & 5.7 & 17 & 8.2 &21 \nl
HR 1729 & \nodata & \nodata & $\ast$ & 376 d & 8.1 & 13 & 5.9 &24 \nl
HR 3951 & \nodata & \nodata & $\ast$ & 6.4 d & 7.6 & 12 & 7.6 &27 \nl
HR 4496 & \nodata & \nodata & $\ast$ & 2.0 d & 12 & 17 & 9.2 & 17 \nl
HR 5019 & \nodata & \nodata & $\ast$ & 9.8 yr & 15 & 17 & 7.6 & 29 \nl
HR 5568 & \nodata & \nodata & $\ast$ & 4.1 d & 9.7 & 11 & 7.1 &40\nl
HR 5868 & \nodata & \nodata & $\ast$ & \nodata & \nodata & \nodata & \nodata &26 \nl
HR 5914 & \nodata & \nodata & $\ast$ & \nodata & \nodata & \nodata & \nodata &15 \nl
HR 7602 & \nodata & \nodata & $\ast$ & 13 yr & 13 & 12 & 5.5 &52 \nl
\nl
\multicolumn{9}{c}{Chromospherically Active Stars ($P_{rot}\leq 14$ days)}\nl
\nl
HR 88 & \nodata &$\ast$ & $\ast$ & \nodata & \nodata & \nodata & \nodata &8 \nl
HR 996 & \nodata & \nodata & $\ast$ & 47 d & 29 & 23 & 16 & 9 \nl
HR 1084 & \nodata & \nodata & $\ast$ & 6.9 yr & 15 & 20 & 13 & 12 \nl
HR 4345 & \nodata & $\ast$ & $\ast$ & 3 d & 34 & 25 & 20 & 8 \nl
HR 4540 & \nodata & \nodata & $\ast$ & 11 d & 15 & 22 & 12 & 14 \nl
HR 4983 & \nodata & \nodata & $\ast$ & 352 d & 25 & 24 & 18 & 12 \nl
HR 5933 & \nodata & \nodata & $\ast$ & \nodata & \nodata & \nodata & \nodata &3 \nl
HR 7061 & \nodata & $\ast$ & $\ast$ & 5.3 d & 94 & 89 & 102 &2 \nl
GL 641 & \nodata & $\ast$ &\nodata& \nodata & \nodata &\nodata & \nodata & 11\nl
\enddata
\tablenotetext{a}{We do not include those stars with confirmed
planetary mass companions (listed in Table \ref{tab:planets}), or
those with stellar or substellar companions (HR 2047, HR 5273, HR
5553, HR 6623 and GL 688; see \S \ref{sec:planets}).}
\tablenotetext{b}{The best fit slope for those stars with significant
trends $>5\ms$ (Table \ref{tab:trends}, \S \ref{sec:trends}).}
\tablenotetext{c}{A ``$\ast$'' indicates a star with significant
variability according to the $\chi^2$ test (Table
\ref{tab:variability}, \S \ref{sec:variability}) or the $\bar{z}$
test (Table \ref{tab:zav}, \S \ref{sec:periodicity}).}
\tablenotetext{d}{For those stars with significant periodicities
(Table \ref{tab:periodogram}, \S \ref{sec:periodicity}), we give
the period and velocity amplitude $K$ of the best fit
sinusoid, $\sigma_{\rm exp}$, the predicted rms taking into account
Doppler errors and intrinsic variability, and $\sigma_{\rm rms}$, the
rms of the residuals to the best fit sinusoid.}
\end{deluxetable}


\begin{deluxetable}{lrrr|llrrr}
\small 
\tablecaption{Upper Limits on the Velocity\tablenotemark{a}\label{tab:K}}
\tablewidth{0pt}
\tablehead{
\colhead{Star} & \colhead{$N$} & \colhead{$\sigma_{\rm rms}$\tablenotemark{b}} &
\colhead{$\bar{K}$} \vline& \colhead{} &
\colhead{Star} & \colhead{$N$} & \colhead{$\sigma_{\rm rms}$\tablenotemark{b}} &
\colhead{$\bar{K}$} \nl
\colhead{(HR)} & \colhead{} & \colhead{(m/s)} &
\colhead{(m/s)} \vline& \colhead{} &
\colhead{(HR)} & \colhead{} & \colhead{(m/s)} &
\colhead{(m/s)} 
}
\startdata
8 & 41 & 18 &25 &&4983 & 100 & 27 &28\nl
88 & 56 & 41 &53 &&5011 & 40 & 27 &38\nl
166 & 70 & 12 &11 &&5019 & 46 & 15 &19\nl
219a & 69 & 19 &10 &&5384 & 29 & 15 &15\nl
222 & 40 & 14 &12 &&5447 & 48 & 51 &73\nl
493 & 38 & 8.9 &10 &&5534 & 24 & 33 &50\nl
509 & 278 & 10 &6 &&5544a & 60 & 34 &45\nl
582 & 34 & 15 &17 &&5544b & 23 & 33 &46\nl
753 & 33 & 11 &9.8 &&5568 & 47 & 11 &15\nl
799 & 43 & 15 &17 &&5868 & 106 & 11 &9.2\nl
857 & 27 & 25 &49 &&5914 & 83 & 16 &12\nl
937 & 131 & 12 &9.4 &&5933 & 70 & 59 &61\nl
962 & 29 & 30 &44 &&6171 & 40 & 11 &11\nl
996 & 67 & 24 &36 &&6458 & 72 & 14 &12\nl
1084 & 121 & 19 &20 &&6806 & 39 & 15 &17\nl
1101 & 40 & 15 &15 &&6869 & 47 & 15 &15\nl
1262 & 30 & 21 &28 &&7061 & 78 & 130 &140\nl
1325 & 66 & 19 &11 &&7462 & 42 & 12 &10\nl
1614 & 30 & 7.7 &7.7 &&7503 & 62 & 19 &14\nl
1729 & 106 & 9.4 &12 &&7602 & 85 & 13 &12\nl
1925 & 12 & 19 &44 &&7672 & 74 & 16 &18\nl
2483 & 61 & 14 &14 &&8085 & 100 & 11 &7.9\nl
2643 & 36 & 20 &16 &&8086 & 58 & 11 &12\nl
3262 & 56 & 22 &26 &&8314 & 36 & 34 &48\nl
3538 & 32 & 16 &18 &&8382 & 21 & 7.8 &9.8\nl
3625 & 28 & 24 &31 &&8665 & 61 & 26 &32\nl
3881 & 62 & 12 &11 &&8832 & 45 & 13 &14\nl
3951 & 75 & 11 &11 &&8969 & 60 & 24 &21\nl
4112 & 35 & 20 &26 &&GL641 & 29 & 24 &44\nl
4345 & 36 & 30 &53 &&GL250a & 18 & 15 &21\nl
4496 & 72 & 15 &15 &&GL716 & 13 & 16 &16\nl
4540 & 80 & 18 &21 &&\nl
\enddata
\tablenotetext{a}{Here we give the mean 99\% upper limit on the
velocity amplitude $\bar{K}$. For each star, we plot this as a dotted line in
Figure \ref{fig:allup}. For most stars, it gives a good estimate of
the upper limit for periods less than the duration of the observations
($\sim$ 11 years for most stars, see Table \ref{tab:stars}).}
\tablenotetext{b}{The rms of the data for each star. For those stars
with a significant long term trend (\S \ref{sec:trends}), the rms is
calculated after subtracting the trend from the data.}
\end{deluxetable}


\begin{deluxetable}{cc} 
\small
\tablecaption{Distribution of Periodogram Powers\label{tab:app}}
\tablewidth{0pt}
\tablehead{
\colhead{Normalization} & \colhead{Prob$(z>z_0)$\tablenotemark{a}} \nl
\colhead{Factor}        &                \nl
}
\startdata
$\sigma^2$ & $\exp(-z_0)$ \nl\nl
$s^2$ & $1-I_{2z_0/(N-m)}\left(1,{N-m-2\over 2}\right)=
\left(1-{2z_0\over N-m}\right)^{N-m-2\over 2}$ \nl\nl
$s_n^2$ & $\int_{z_0}^{\infty}dz\ F_{2,N-m-2}(z)=
\left(1+{2z_0\over N-m-2}\right)^{-{N-m-2\over 2}}$ \nl\nl
\enddata
\tablenotetext{a}{$I_x(a,b)$ is the incomplete beta function
(Abromowitz \& Stegun 1971); $F_{\nu_1,\nu_2}$ is Fisher's $F$
distribution with $\nu_1$ and $\nu_2$ degrees of freedom (Hoel, Port
\& Stone 1971).}
\end{deluxetable}


\clearpage

\begin{figure}
\epsscale{1.0}
\plotone{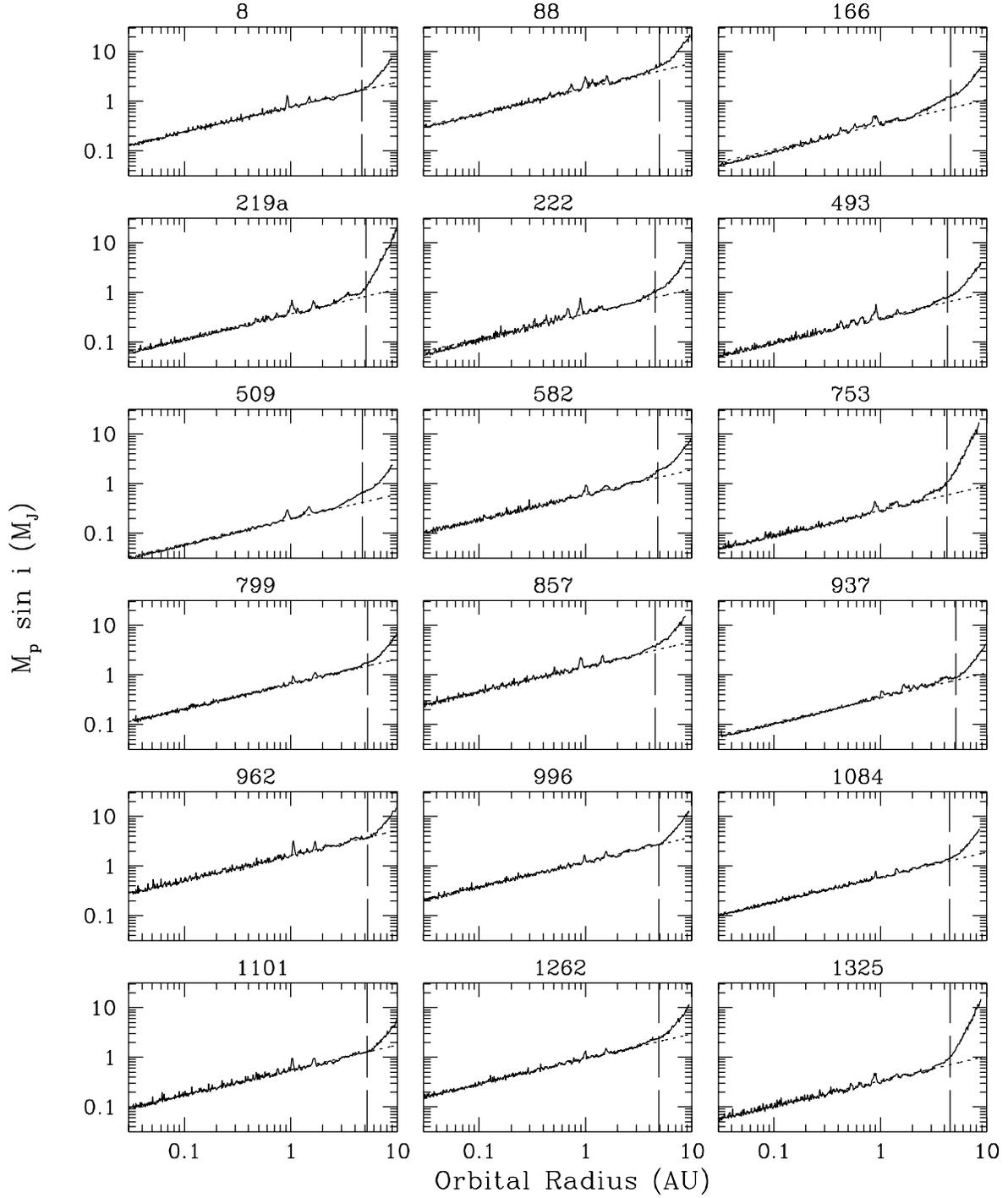}
\caption{ The 99\% upper limit on the mass ($\mpsini$) of a
companion in a circular orbit as a function of orbital radius for each
of the sample stars. The upper limit is calculated as described in \S
\ref{sec:method}. The dotted line shows a line of constant velocity,
where the velocity for each star is the mean upper limit given in
Table \ref{tab:K}. The long dashed line shows the orbital radius at
which the duration of the observations equals the orbital period.
\label{fig:allup}}
\end{figure}

\begin{figure}
\figurenum{\ref{fig:allup}}
\plotone{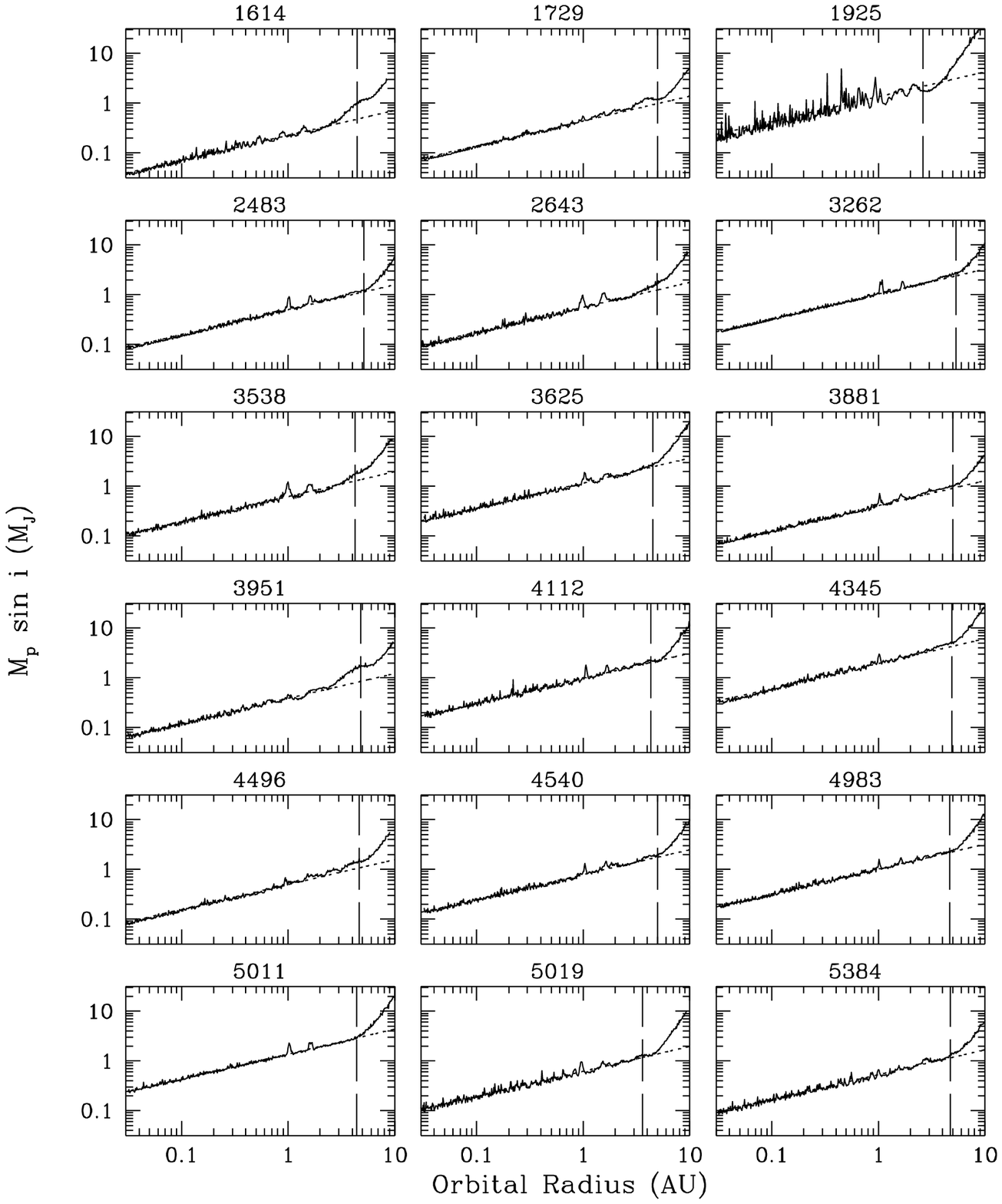}
\caption{{\it Continued.}}
\end{figure}

\begin{figure}
\figurenum{\ref{fig:allup}}
\plotone{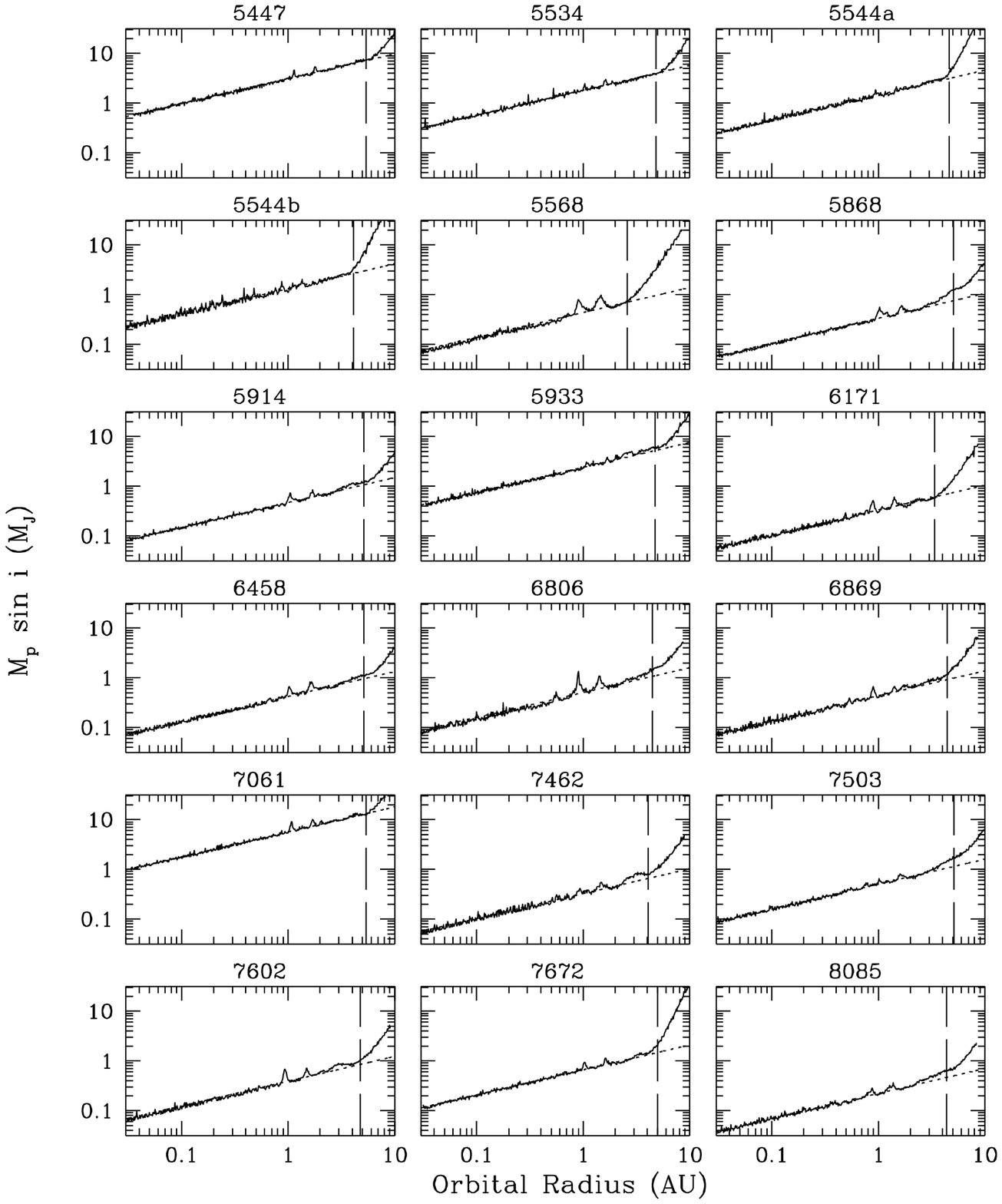}
\caption{{\it Continued.}}
\end{figure}

\begin{figure}
\figurenum{\ref{fig:allup}}
\plotone{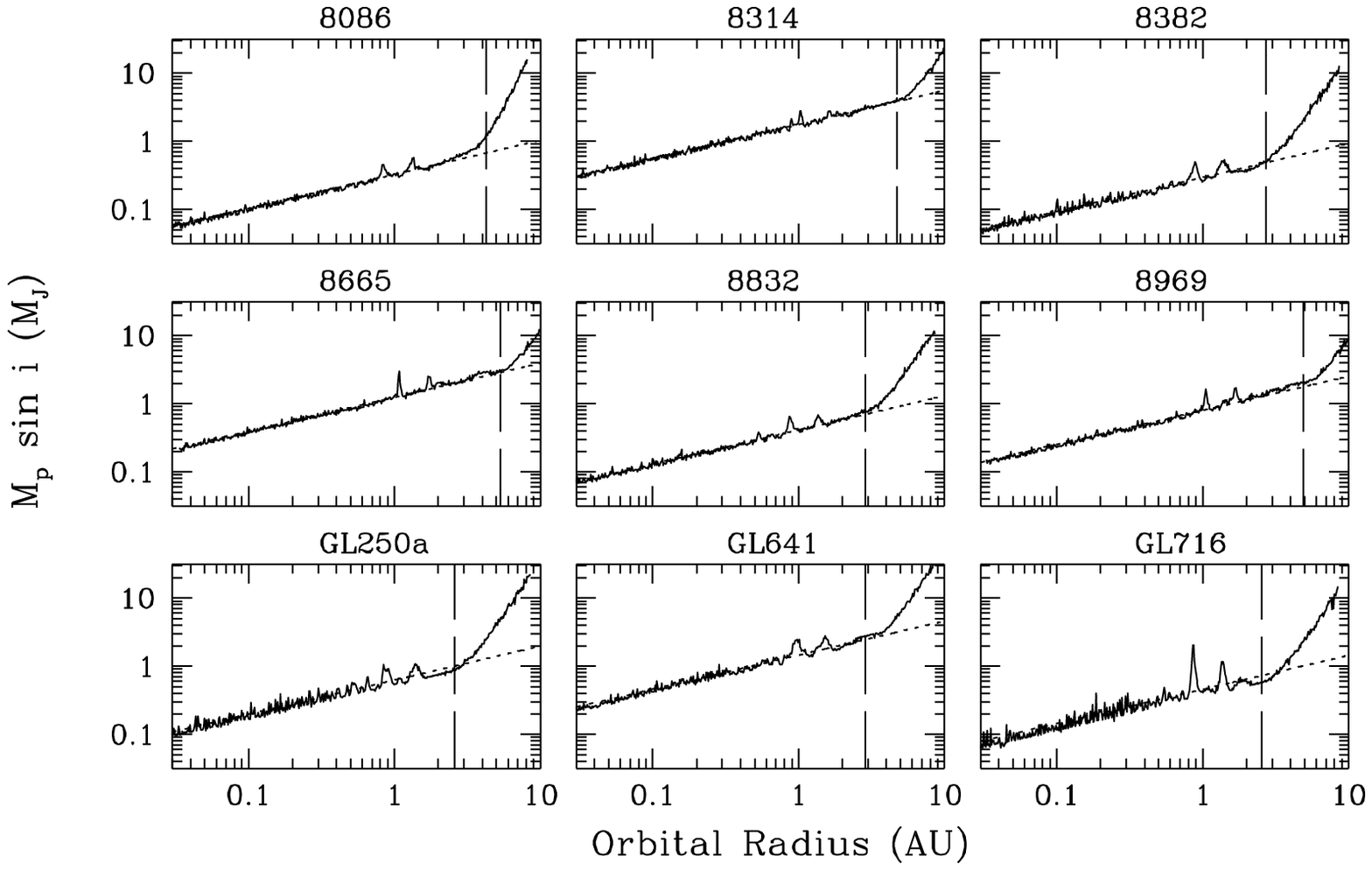}
\caption{{\it Continued.}}
\end{figure}

\begin{figure}
\epsscale{0.6}
\plotone{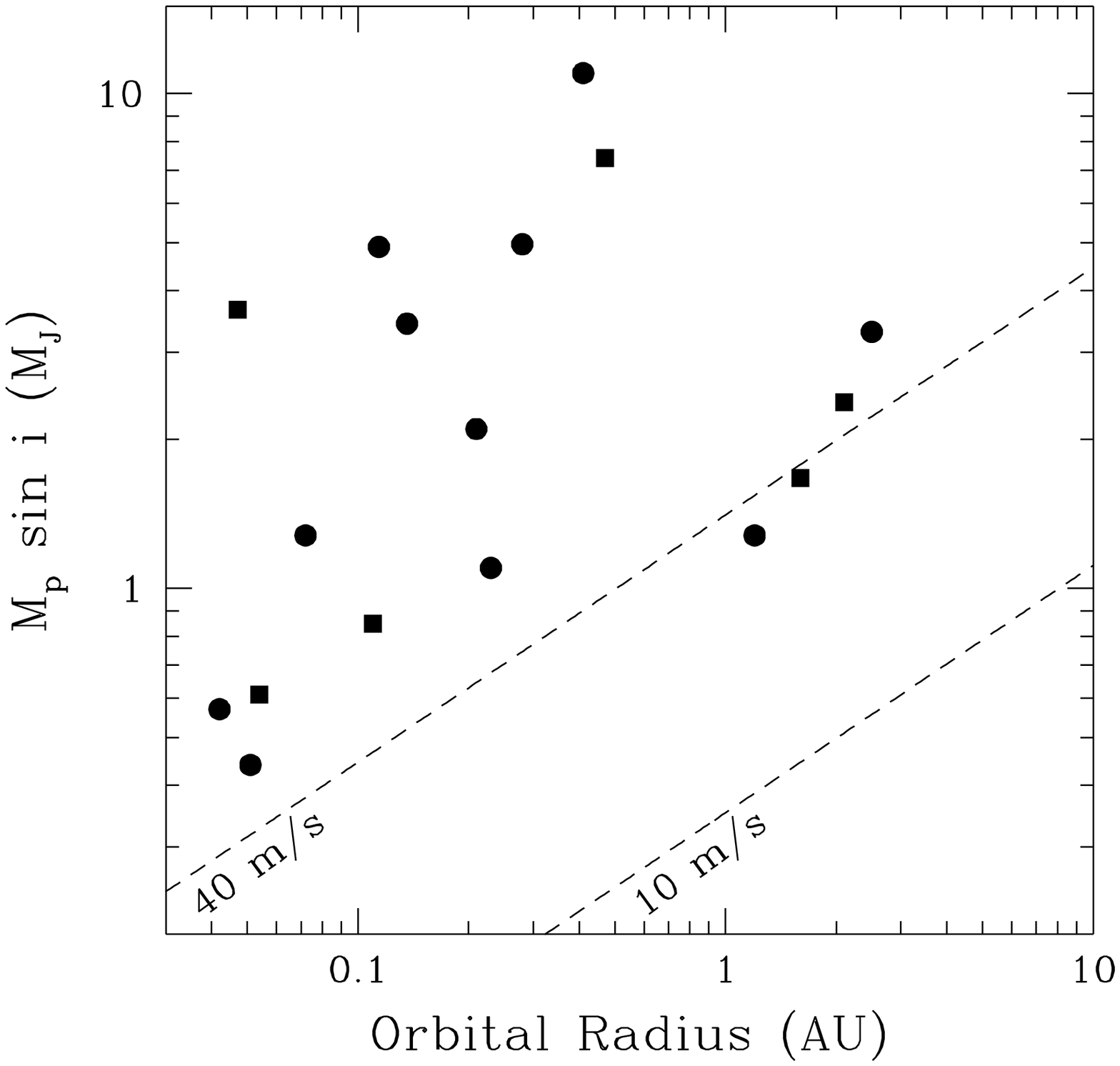}
\caption{ 
The mass ($M_p\sin i$) and orbital radius (semi-major axis for
eccentric orbits) of confirmed companions with $\mpsini<15 M_J$. The
orbital parameters and references are given in Table
\ref{tab:planets}. Companions discovered at Lick Observatory and
included in our sample of stars are shown as solid squares. The dashed
lines are lines of constant velocity amplitude for a star with $M=1
M_\odot$.
\label{fig:planets}}
\end{figure}

\begin{figure}
\epsscale{1.0}
\plotone{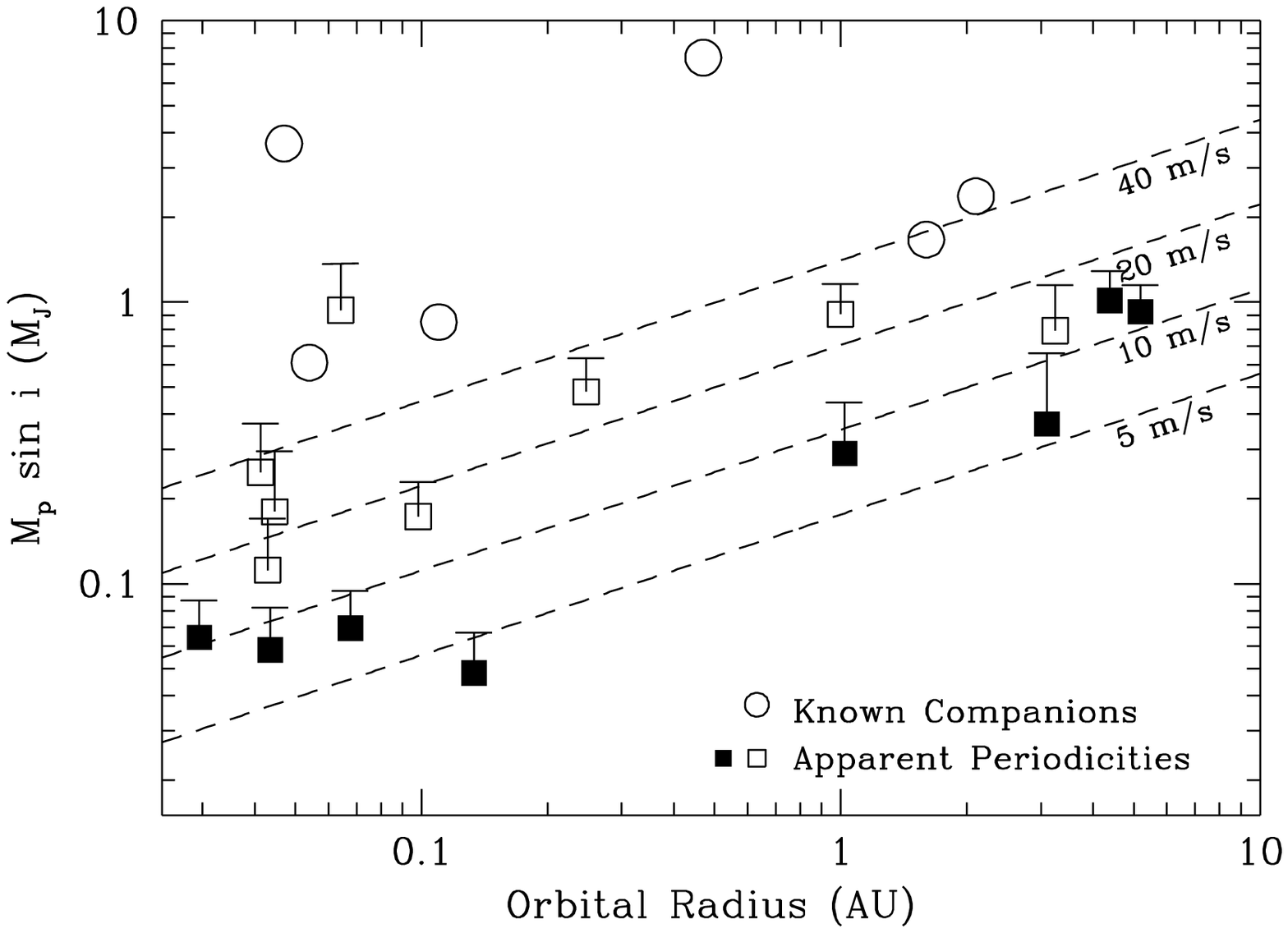}
\caption{Confirmed companions and apparent periodicities from our
search for companions, plotted in the $\mpsini$--$a$ plane. We plot
the six confirmed planetary-mass companions in our sample as
circles. The squares show the other significant periodicities revealed
by our periodogram analysis (Table \ref{tab:summary}), presented in
this paper for the first time. These periodicities are {\it as yet
unconfirmed} candidates for companions. Open squares indicate
chromospherically active stars ($\prot\leq 14$ days); filled squares
indicate chromospherically quiet stars ($\prot>14$ days). The errorbar
on each point shows the 99\% upper limit on the velocity amplitude
calculated as in \S \ref{sec:method}. The dashed lines show lines of
constant velocity for a solar mass star.
\label{fig:candid}}
\end{figure}

\begin{figure}
\epsscale{1.0}
\plotone{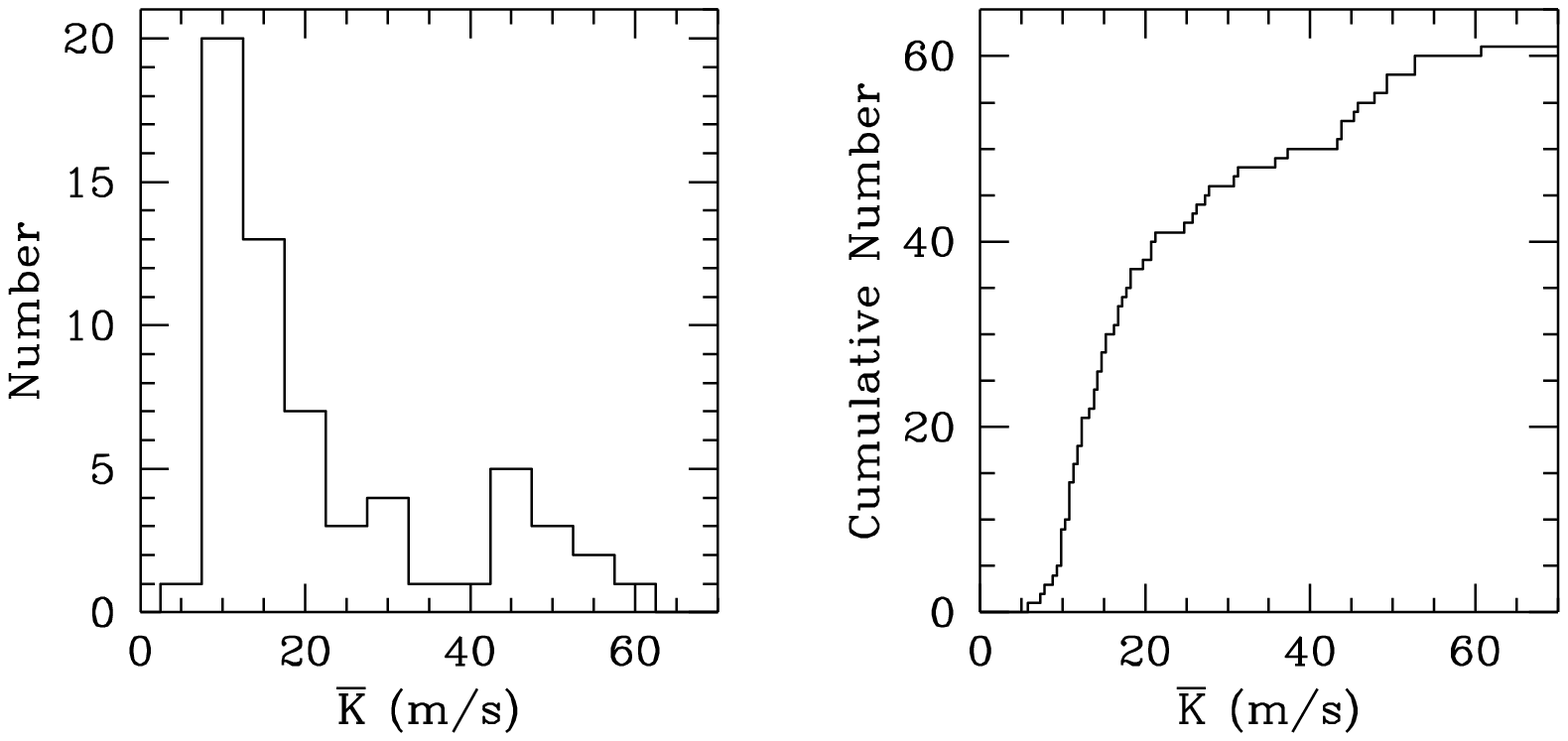}
\caption{ The distribution of the mean 99\% upper limits on velocity
amplitude listed in Table \ref{tab:K}. The left panel shows a
histogram of the mean upper limits (in $5\ms$ bins; two stars have
$\bar{K}>70\ms$). The right panel shows the cumulative distribution of
mean upper limits. The total number of stars is 63.
\label{fig:kdist}}
\end{figure}

\begin{figure}
\epsscale{1.0}
\plotone{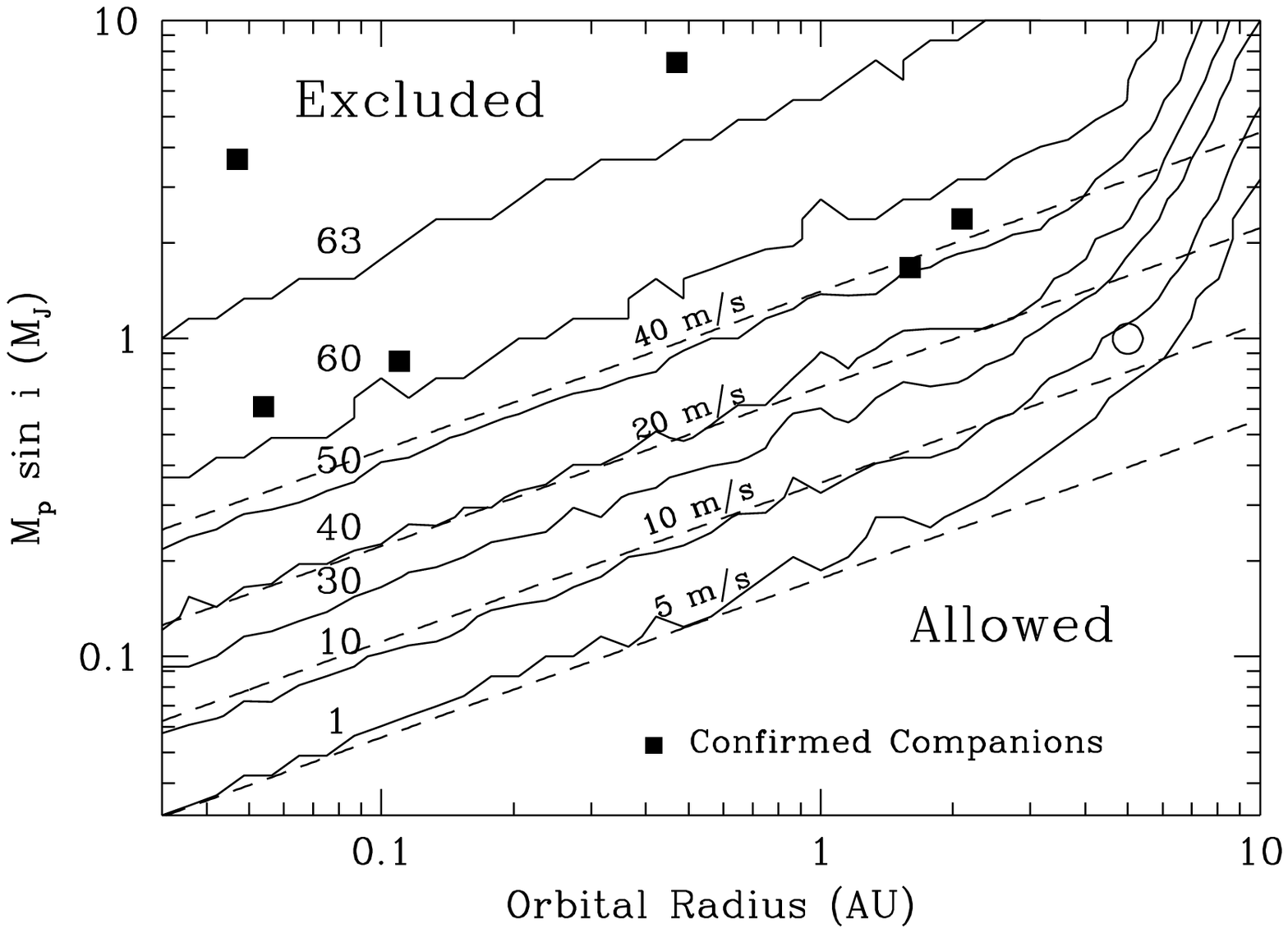}
\caption{ Contours of the number of stars for which a companion of
particular $\mpsini$ and $a$ can be excluded at the 99\% level, for
the set of 63 stars shown in Figure \ref{fig:allup}. The contours
(solid lines) are plotted on a $40\times 40$ grid and are labelled
with the number of stars excluded from the region up and to the left
of the line. We plot as squares those confirmed companions (Table
\ref{tab:planets}) discovered at Lick. Dashed lines give $5, 10, 20$
and $40\ms$ constant velocity lines for a $1 M_\odot$ star. This
figure is a two-dimensional version of the right panel of Figure
\ref{fig:kdist}. The circle shows $\mpsini=1\mj$, $a=5.2\au$.
\label{fig:ncont}}
\end{figure}

\begin{figure}
\epsscale{1.0}
\plottwo{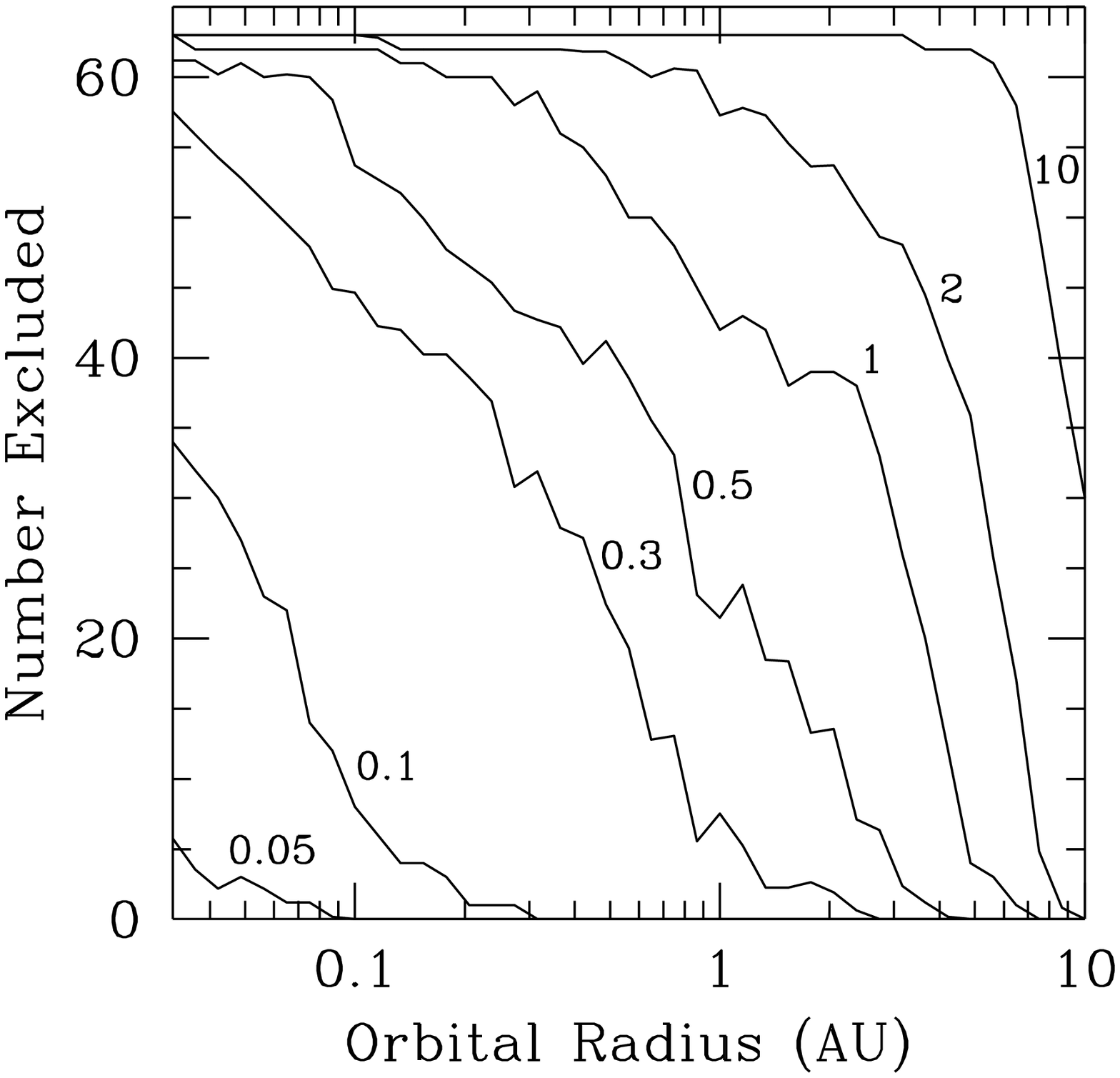}{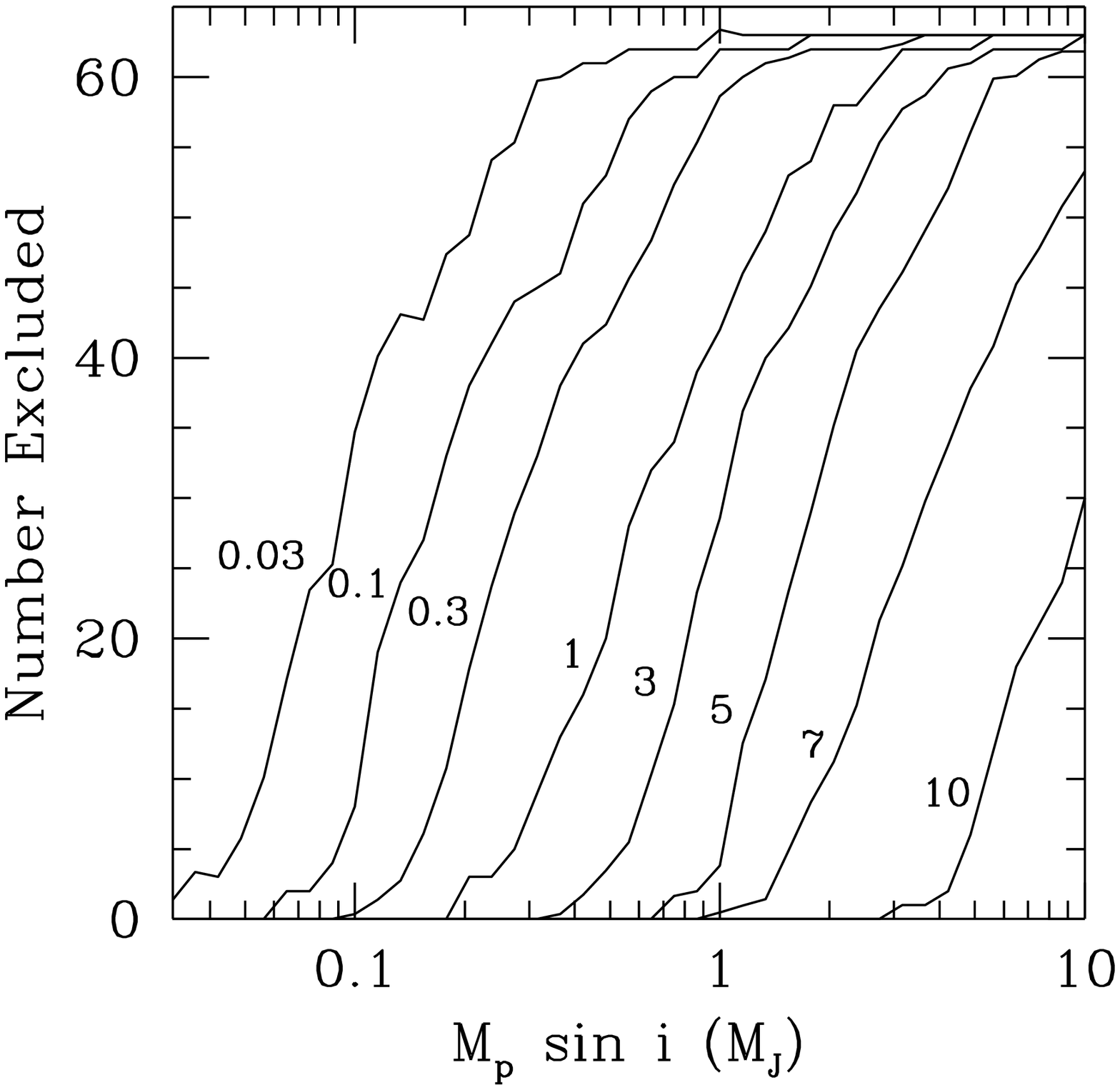}
\caption{ 
The number of stars from which a companion with a particular $a$ and
$\mpsini$ can be excluded at the 99\% level. We show the number
excluded as a function of $a$ for different $\mpsini$ (left panel) and
as a function of $\mpsini$ for different $a$ (right panel). Each curve is 
a different section of the contour map of Figure
\ref{fig:ncont}. The curves are labelled with the corresponding values
of $a$ or $\mpsini$.
\label{fig:nxb}}
\end{figure}

\begin{figure}
\epsscale{0.6}
\plotone{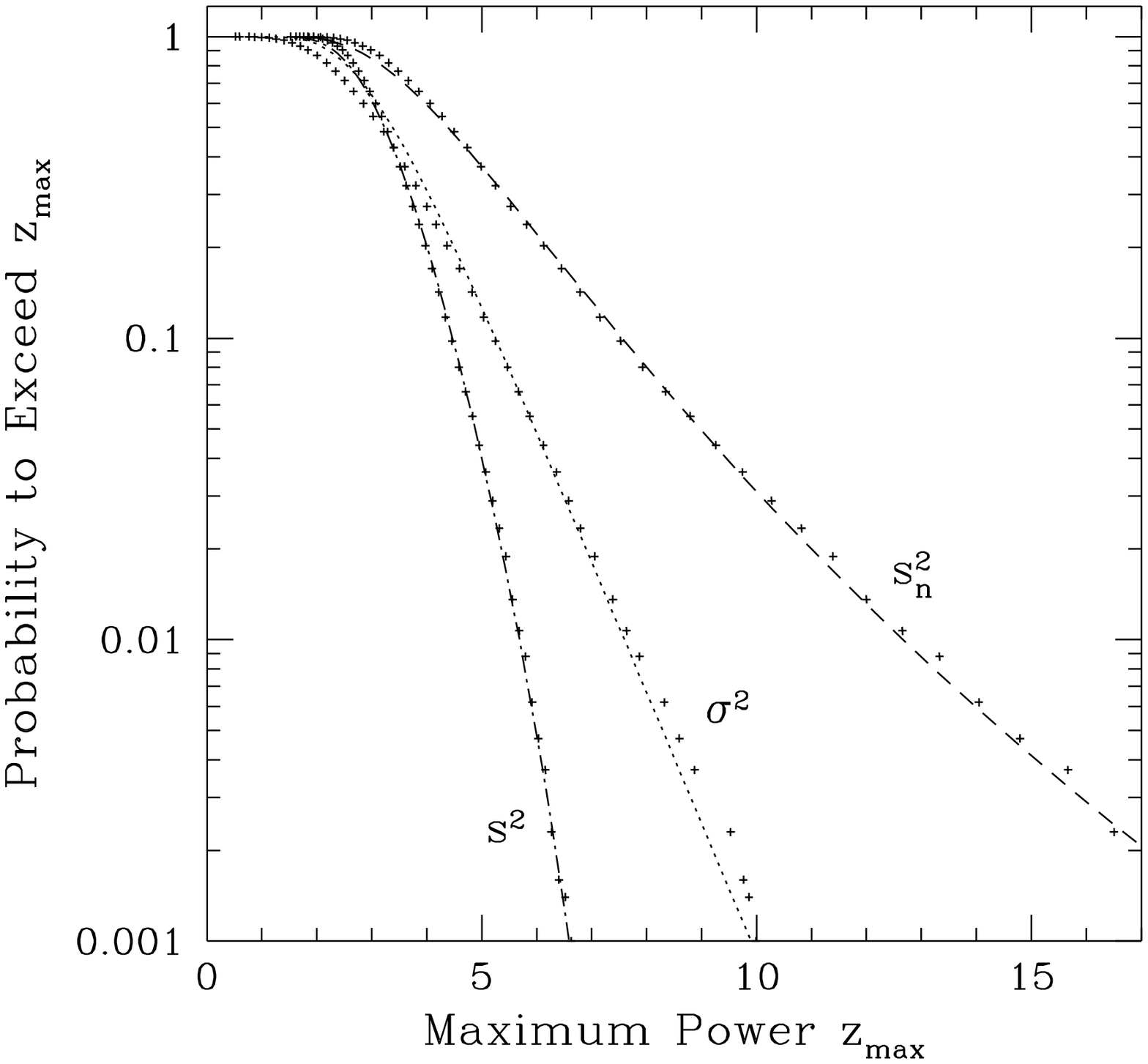}
\caption{
The distribution of maximum periodogram powers with
different normalizations. The crosses are the results of Monte Carlo
simulations. We generate ten thousand sets of $N=20$ evenly spaced
points drawn from a Gaussian distribution. The periodogram is
evaluated at $2N$ evenly spaced frequencies between $1/N$ and
$1/2$. Each curve is labelled with the normalization factor, either
the known variance of the noise $\sigma^2$, the sample variance $s^2$
or the variance of the residuals $s_n^2$. Theoretical distributions
are plotted as lines. The dotted line is the distribution
$1-(1-\exp(-z))^N$. The dashed and dot-dashed lines are of the form
$1-(1-f(z))^M$, where $f(z)$ is taken from Table \ref{tab:app}. The best
fit $M$ was determined by fitting the tail of the $F$ distribution
(probability from 0.5 to 0.99), giving $M=23.5$.
\label{fig:dist}}
\end{figure}

\end{document}